\documentclass[a4paper,twocolumn,superscriptaddress,11pt,accepted=2021-09-17]{quantumarticle}
\pdfoutput=1
\usepackage[utf8]{inputenc}
\usepackage[T1]{fontenc}
\usepackage[english]{babel}
\usepackage{graphicx}  
\usepackage{dcolumn}   
\usepackage{bm}        
\usepackage{amssymb}   
\usepackage{amsmath}
\usepackage{mathtools}
\usepackage[utf8]{inputenc}
 \usepackage[numbers, compress]{natbib}
\usepackage{changes}

\definecolor{myurlcolor}{rgb}{0,0,0.7}
\definecolor{myrefcolor}{rgb}{0.8,0,0}
\usepackage{hyperref}
\hypersetup{colorlinks, linkcolor=myrefcolor,
citecolor=myurlcolor, urlcolor=myurlcolor}

\usepackage{cleveref}
\crefrangeformat{equation}{#3(#1 #4--#5 #2)#6}

\usepackage{comment}
\usepackage{amsthm}

\newcommand{\ignore}[1]{}

 \definecolor{jordi}{rgb}{0.1,0.1,0.5}

\usepackage[T1]{fontenc}
\usepackage{graphicx}
\usepackage{amssymb}
\usepackage{amsmath}
\usepackage{color}
\usepackage{psfrag}
\usepackage{epsfig}
\usepackage{bbm}
\usepackage{bm}
\usepackage{hyperref}
\usepackage[normalem]{ulem}
\usepackage{amsthm}
\usepackage{epstopdf}
\usepackage{graphicx}
\usepackage{stackengine,xcolor}
\usepackage{tcolorbox}

\definecolor{nred}{rgb}{0.9,0.1,0.1}
\definecolor{nblack}{rgb}{0,0,0}
\definecolor{nblue}{rgb}{0.2,0.2,0.8}
\definecolor{ngreen}{rgb}{0.2,0.6,0.2}
\definecolor{nmagenta}{rgb}{0.54,0,0.54}

\newcommand{\blu}{\color{nblack}}
\newcommand{\blub}{\color{nblack}}
\newcommand{\blubb}{\color{nblack}}
\newcommand{\bluiv}{\color{nblack}}
\newcommand{\bluv}{\color{nblack}}

\usepackage{etoolbox}
\usepackage{bbold} 

\newcommand{\ket}[1]{| #1 \rangle}
\newcommand{\bra}[1]{\langle #1 |}

\newcommand{\beq}{\begin{eqnarray}}
\newcommand{\eeq}{\end{eqnarray}}

\newcommand{\rab}{{\varrho^{\text{AB}}}}
\newcommand{\rca}{{\varrho^{\text{CA}_0}}}
\newcommand{\rbd}{{\varrho^{\text{B}_0\text{D}}}}
\newcommand{\rax}{{\rho_{a|x}}}
\newcommand{\hatrax}{{\hat{\rho}_{a|x}}}
\newcommand{\sax}{{\sigma_{a|x}}}
\newcommand{\hatsax}{{\hat{\sigma}_{a|x}}}
\newcommand{\hatsaxstar}{{\hat{\sigma}^*_{a|x}}}

\newcommand{\tcu}{{\tau_{c|u}}}
\newcommand{\hattcu}{{\hat{\tau}_{c|u}}}
\newcommand{\tcustar}{{\tau^*_{c|u}}}
\newcommand{\tcustarstar}{{\tau^{**}_{c|u}}}
\newcommand{\hattcustar}{{\hat{\tau}^*_{c|u}}}

\newcommand{\odv}{{\omega_{d|v}}}
\newcommand{\hatodv}{{\hat{\omega}_{d|v}}}
\newcommand{\odvstar}{{\omega^*_{d|v}}}
\newcommand{\odvstarstar}{{\omega^{**}_{d|v}}}
\newcommand{\hatodvstar}{{\hat{\omega}^*_{d|v}}}

\newcommand{\Dcu}{{\Delta_{c|u}}}
\newcommand{\Ddv}{{\Delta_{d|v}}}

\newcommand{\sigmakA}{{\hat{\sigma}^{\text{A}}_k}}
\newcommand{\sigmakB}{{\hat{\sigma}^{\text{B}}_k}}

\newcommand{\psax}{{P_{\bm{\sigma}}(a|x)}}
\newcommand{\psaxstar}{{P_{\bm{\sigma}}^*(a|x)}}
\newcommand{\psaxstarstar}{{P_{\bm{\sigma}}^{**}(a|x)}}

\newcommand{\prax}{{P_{\bm{\rho}}(a|x)}}
\newcommand{\Eax}{{E^\text{A}_{a|x}}}
\newcommand{\Eby}{{E^\text{B}_{b|y}}}
\newcommand{\EAA}{{E^{\text{A}_0\text{A}}}}
\newcommand{\EBB}{{E^{\text{B}\text{B}_0}}}

\DeclareMathOperator{\tr}{tr}

\theoremstyle{definition}
\newtheorem{definition}{Definition}

\begin{document}
\selectlanguage{english}

\title{Robust self-testing of steerable quantum assemblages and its applications on device-independent quantum certification}

\author{Shin-Liang Chen}
\email{shin.liang.chen@phys.ncku.edu.tw}
\affiliation{Department of Physics and Center for Quantum Frontiers of Research \& Technology (QFort), National Cheng Kung University, Tainan 701, Taiwan}
\affiliation{Dahlem Center for Complex Quantum Systems, Freie Universit\"at Berlin, 14195 Berlin, Germany}
\affiliation{Max-Planck-Institut f{\"u}r Quantenoptik, Hans-Kopfermann-Stra{\ss}e 1, 85748 Garching, Germany}
\orcid{0000-0002-3453-4794}

\author{Huan-Yu Ku}
\email{huan{\_}yu@phys.ncku.edu.tw }
\affiliation{Department of Physics and Center for Quantum Frontiers of Research \& Technology (QFort), National Cheng Kung University, Tainan 701, Taiwan}
\orcid{0000-0003-1909-6703}

\author{Wenbin Zhou}
\affiliation{Graduate School of Informatics, Nagoya University, Chikusa-ku, 464-8601 Nagoya, Japan}
\orcid{0000-0002-8213-1578}

\author{Jordi Tura}
\affiliation{Max-Planck-Institut f{\"u}r Quantenoptik, Hans-Kopfermann-Stra{\ss}e 1, 85748 Garching, Germany}
\affiliation{Instituut-Lorentz, Universiteit Leiden, P.O. Box 9506, 2300 RA Leiden, The Netherlands}
\orcid{0000-0002-6123-1422}

\author{Yueh-Nan Chen}
\email{yuehnan@mail.ncku.edu.tw }
\affiliation{Department of Physics and Center for Quantum Frontiers of Research \& Technology (QFort), National Cheng Kung University, Tainan 701, Taiwan}
\orcid{0000-0002-2785-7675}

\maketitle

\begin{abstract}
Given a Bell inequality, if its maximal quantum violation can be achieved only by a single set of measurements for each party or a single quantum state, up to local unitaries, one refers to such a phenomenon as \emph{self-testing}. For instance, the maximal quantum violation of the Clauser-Horne-Shimony-Holt (CHSH) inequality certifies that the underlying state contains the two-qubit maximally entangled state and the measurements of one party (say, Alice) contains a pair of anti-commuting qubit observables. As a consequence, the other party (say, Bob) automatically verifies his set of states remotely steered by Alice, namely the \emph{assemblage}, is in the eigenstates of a pair of anti-commuting observables. It is natural to ask if the quantum violation of the Bell inequality is not maximally achieved, or if one does not care about self-testing the state or measurements, are we capable of estimating how close the underlying assemblage is to the reference one? In this work, we provide a systematic device-independent estimation by proposing a framework called \emph{robust self-testing of steerable quantum assemblages}. In particular, we consider assemblages violating several paradigmatic Bell inequalities and obtain the robust self-testing statement for each scenario. Our result is device-independent (DI), i.e., no assumption is made on the shared state and the measurement devices involved. Our work thus not only paves a way for exploring the connection between the boundary of quantum set of correlations and steerable assemblages, but also provides a useful tool in the areas of device-independent quantum certification. As two explicit applications, we show 1) that it can be used for an alternative proof of the protocol of DI certification of all entangled two-qubit states proposed by Bowles \emph{et al.} [Phys. Rev. Lett. 121, 180503 (2018)], and 2) that it can be used to verify all non-entanglement-breaking qubit channels with fewer assumptions compared with the work of Rosset \emph{et al.} [Phys. Rev. X 8, 021033 (2018)].
\end{abstract}

Nonlocality of quantum theory enables one, by performing incompatible measurements on entangled states, to create correlations not admitting a local-hidden-variable model~\cite{Bell64}. Such correlations, termed \emph{nonlocal correlations}, can be observed by violating a Bell inequality~\cite{Bell64,Freedman72,Brunner14} and allow one to perform quantum certification tasks in a device-independent (DI) way~\cite{Acin07,Scarani12,Brunner14}, in the sense that one makes no assumption on the measurement devices or the shared quantum states. For instance, observing a Bell inequality violation verifies, in a DI manner, that the shared state is entangled~\cite{Wiseman07,Quintino15} and that the measurements performed are incompatible~\cite{Wolf09}. Remarkably, in the extreme case, such as one obtains the maximal quantum violation of certain Bell inequalities, one is able to verify the exact quantum description of the state and measurements. For instance, observing the maximal quantum violation of the Clauser-Horne-Shimony-Holt (CHSH) inequality~\cite{Clauser69} uniquely certifies that the system under consideration contains the maximally entangled two-qubit state and that a pair of anti-commuting qubit observables is embedded in the measurements performed~\cite{Mayers98,Mayers04}. Since then, such a kind of certification, dubbed as \emph{self-testing}, has been used for verifying various of quantum states and measurements with distinct Bell inequalitites (see Ref.~\cite{Supic19} for a review). Importantly, if one is still capable of estimating how close the underlying system is to the ideal system even when the violation of the Bell inequality departs from the maximal quantum value, then the self-testing is \emph{robust}, which is an essential property for both practical point of view and experimental demonstrations.

\begin{figure*}
\centering
\includegraphics[width=0.7\linewidth]{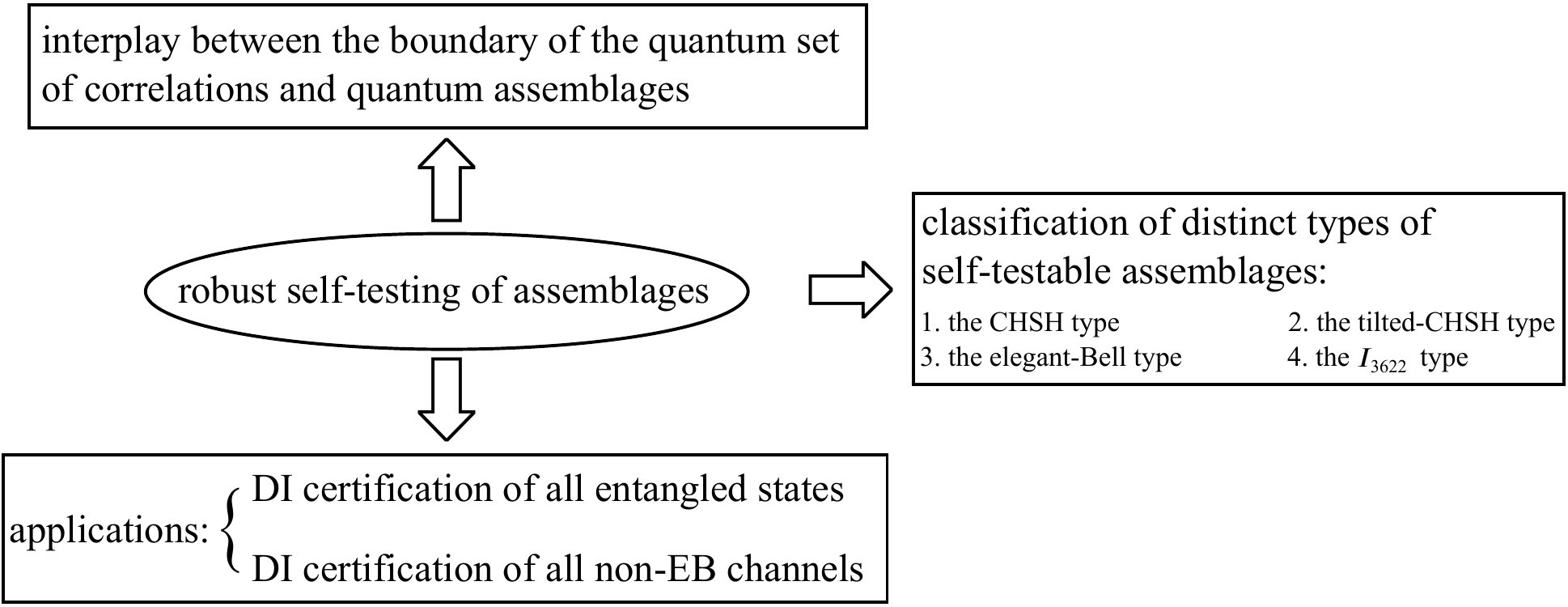}
\caption{The overview of this work.
}
\label{Fig_picture}
\end{figure*}

Apart from nonlocality, another intriguing phenomenon occurring between spatially separated systems is \emph{steering}~\cite{Schrodinger35,Wiseman07,Cavalcanti17,Uola2020}. Consider two parties, called Alice and Bob, sharing a quantum state. By locally performing incompatible measurements on her part of share, Alice remotely steers Bob's share into a set of states with certain probabilities. The set of such states and probabilities, referred to as the \emph{assemblage}~\cite{Pusey13}, is the resource quantity concerned in a steering-type experiment and plays an important role in the resource theory of steering~\cite{Pusey13,SNC14,Piani15,Gallego15,HLL2016,KuPRA18}. Operationally, quantum steerability can be treated as an entanglement verification task in an asymmetric quantum network~\cite{Wiseman07,Cavalcanti15}, crucial to demonstration of quantum key distribution in practice~\cite{Branciard12,Gallego15}. As applications in quantum cryptography, it was shown that quantum steerability can be used for the optimal randomness certification~\cite{Passaro15} and maximal randomness expansion~\cite{Cavalcanti18}. It was also found that steerability is closely relalted to measurement incompatibility~\cite{Quint14,Uola14}, hence being an essential bridge to study measurement incompatibility~\cite{Uola15,Cavalcanti16,CBLC16}. All of the above pieces of research, as well as many other works related to steering, rely on the analysis of assemblages. Therefore, studying the property of assemblages allows one to understand steering more deeply and gives a further boost to applications of quantum information processing.

In this work, we propose a method for verifying the assemblage when given a Bell inequality violation. For instance, we show that when observing the maximal quantum violation of the CHSH inequality, Bob's assemblage must contain the set of eigenstates of anti-commuting observables, yielding equivalent probabilities. This seems not surprising since the maximal violation has already told us that the shared state contains the maximally two-qubit entangled state and Alice's measurements form an anti-commuting set of observables, therefore the information of Bob's assemblage can be obtained by the rule of quantum theory. What we are mainly concerned with is the \emph{imperfect} situation, that is, our proposed method is capable of estimating how close the underlying assemblage is to the ideal one when the violation of the CHSH inequality departs from its maximal value. We refer to the method as \emph{robust self-testing of quantum assemblages}.
{\blub There are two main ideas behind our framework. First, the definition of self-testing considered in this work is based on the formulation of \emph{extractability} proposed by Kaniewski~\cite{Kaniewski16}. That is, the goal is to search for a completely positive and trace-preserving map which maps the underlying assemblage to the reference assemblage. Second, the concept behind our numerical computation is borrowed from the idea of the swap method~\cite{Yang14,Bancal15} ---relaxing the ideal case to a DI setting.}
Naturally, while the swap method uses the typical numerical approximation, i.e., the semidefinite relaxation~\cite{Doherty08,NPA,NPA2008,Moroder13}, to carry out the computation, we employ a variant tool called the \emph{assemblage moment matrices}~\cite{CBLC16,CBLC18}. Apart from the CHSH scenario, we also consider assemblages which maximally violate the tilted CHSH inequality~\cite{Acin12}, the elegant Bell inequality~\cite{Gisin:ManyQuestions}, and the one of Ref.~\cite{Acin16}, and obtain the robust self-testing statement in each scenario.
{\blu Note that in each of Bell scenarios considered in our work, one can also robustly self-test the reference state and measurements through the standard method of self-testing. Therefore, it is natural that one may robustly self-test the assemblage through the idea of error propagation, i.e., the robustness propagates from the state \& measurements to the assemblage. From this point of view, we would like to point out that the robustness given by the standard self-testing is strictly low, while the robustness obtained through our method is much better due to the similar concept of the swap method~\cite{Yang14,Bancal15}. Except for this benefit, we note that different states and measurements (even not unitary equivalent) can generate the same assemblage, therefore it leads to an open question: Does there exist a Bell scheme where one can self-test the assemblage while self-testing the state or measurements is impossible? Thus, another goal of this work is to develop a universal framework for self-testing the assemblages, no matter one can self-test the states or the measurements. }
Fundamentally, our work classifies various types of steerable assemblages and has a deep connection between these assemblages and the boundary of the quantum set of correlations. Our work thus has direct applications on DI quantum certification. Two explicit applications are provided: 1) It can be used for an alternative proof of the DI certification protocol of all entangled two-qubit states proposed in Refs.~\cite{Bowles18,Bowles18PRA}. More explicitly, {\blu while the original proof is based on the protocols of self-testing of the state and the measurements, our proof is based on self-testing of the assemblages.} 2) It can be used for making the verification protocol of all non-entanglement-breaking qubit channels~\cite{Rosset18} fully DI. MATLAB codes to accompany some of the results in this work can be found in Ref.~\cite{SLChen_code}. The general picture of this work is depicted in Fig.~\ref{Fig_picture}.

The rest of paper is organized as follows: In Sec.~\ref{Sec_pre}, we briefly review the concepts of Bell nonlocality and steerability and introduce the notations used in the entire manuscript. In Sec.~\ref{Sec_robust_ST}, we introduce the framework for robust self-testing of assemblages and use the CHSH scenario as a typical example. After that, robust self-testing of assemblages in other paradigmatic Bell scenarios are also explored. In Sec.~\ref{Sec_applications}, we provide two explicit applications on DI certification of entangled states and non-entanglement-breaking channels. In Sec.~\ref{Sec_discussion}, we conclude our work and discuss some possible issues for future research. Some related proofs are included in Appendices.

\section{Preliminaries}\label{Sec_pre}
\subsection{Bell scenario}\label{Sec_Bell}
Let us start by briefly reviewing a Bell-type experiment. Consider a bipartite physical resource shared between two observers, called Alice and Bob. During each round, Alice (Bob) performs a measurement labelled by $x\in\mathcal{X}=\{1,2,3,...,|\mathcal{X}|\}$ ($y\in\mathcal{Y}=\{1,2,3,...,|\mathcal{Y}|\}$) on her (his) part of system and obtains a measurement outcome $a\in\mathcal{A}=\{1,2,3,...,|\mathcal{A}|\}$ ($b\in\mathcal{B}=\{1,2,3,...,|\mathcal{B}|\}$), where $|\mathcal{X}|$ denotes the cardinality of the set $\mathcal{X}$. After many rounds, they observe a set of joint probabilities of measurement outcomes conditional on the measurement settings, namely a \emph{correlation} $\{P(a,b|x,y)\}_{a,b,x,y}:=\mathbf{P}$, which can be treated as a single point in $\mathbb{R}^{|\mathcal{A}||\mathcal{B}||\mathcal{X}||\mathcal{Y}|}$. Apart from the axiomatic constraints on the correlation such as the normalization ($\sum_{a,b}P(a,b|x,y)=1$)\footnote{In the entire paper, for simplicity, we omit the statement of ``for all indices'' such as ``for all $x,y$'' or ``for all $a,b,x,y$'' when there is no risk of confusion.} and the positivity ($P(a,b|x,y)\geq 0$), the correlation $\mathbf{P}$ also suffers from distinct types of constraints in different theories. For instance, if $\mathbf{P}$ is generated by a local-hidden-variable model, the probabilities conditional on the hidden variables $\lambda$ factorize as $P(a,b|x,y,\lambda)= P(a|x,\lambda) P(b|y,\lambda)$. Such correlations are referred to as \emph{local correlations} and their set, denoted by $\mathcal{L}$, forms a \emph{polytope} in $\mathbb{R}^{|\mathcal{A}||\mathcal{B}||\mathcal{X}||\mathcal{Y}|}$, i.e., it is a bounded, convex set with finite number of extremal points. If the correlation is generated by quantum theory, {\blu i.e., $\mathbf{P}$ is generated by Alice and Bob locally performing quantum measurements on the shared bipartite quantum state $\rab$, then the correlation is dubbed as \emph{quantum correlation}. In such a case, $\mathbf{P}$ follows the Born rule: $P(a,b|x,y)=\tr(\Eax\otimes\Eby~\rab)$, where $\{\Eax\}_{a,x}$ and $\{\Eby\}_{b,y}$ are, respectively, positive-operator valued measures (POVMs) which describes Alice's and Bob's quantum measurements.} The set of quantum correlations, denoted by $\mathcal{Q}$, also forms a convex set, but it is not a polytope since it has infinite extreme points. Importantly, $\mathcal{Q}$ is a proper superset of $\mathcal{L}$~\cite{Bell64,Brunner14}. This means for any given quantum correlation $\mathbf{P}'$ which is not local, there exists a hyperplane $\sum_{a,b,x,y}\beta_{a,b,x,y}P(a,b|x,y)=\alpha^L$ separating $\mathbf{P}'$ and $\mathcal{L}$, with $\beta_{a,b,x,y}$ being some real numbers. Consequently, all local correlations must satisfy the so-called {\emph{Bell inequality} $\sum_{a,b,x,y}\beta_{a,b,x,y}P(a,b|x,y)\leq \alpha^L$. Denoting by $\alpha^Q$ the maximal value that the quantity $\sum_{a,b,x,y}\beta_{a,b,x,y}P(a,b|x,y)$ can achieve by quantum correlations~\cite{Tsirelson1980}, then one has $\alpha^L\leq\alpha^Q$ due to $\mathcal{L}\subset\mathcal{Q}$. All the Bell inequalities considered in this work are those with $\alpha^Q$ being strictly greater than $\alpha^L$.

\subsection{Quantum steerability}\label{Sec_steering}
The typical steering scenario~\cite{Wiseman07,SNC14,Piani15,Gallego15} is similar to the Bell scheme. The difference is that during each round of Alice's action $(a,x)$, Bob is able to characterize his part of the quantum state, denoted by $\hatrax$. After many rounds, Bob obtains a collection of ensembles $\{\hatrax\}_{a,x}$ remotely prepared by Alice, whose measurement statistics are described by a set of conditional probabilities $\{P(a|x)\}_{a,x}$. It is convenient to introduce a quantity, dubbed as an \emph{assemblage}~\cite{Pusey13,SNC14,Piani15,Gallego15}, defined as $\{\rax:=P(a|x)\hatrax\}_{a,x}$ so that Alice's statistics and Bob's collections of ensembles can be both characterized through the relations: $P(a|x)=\tr(\rax)$ and $\hatrax=\rax/\tr(\rax)$. Compared to a local-hidden-variable model in the Bell scenario, the classical counterpart in the steering scenario is the so-called \emph{local-hidden-state (LHS) model}. That is, the bipartite physical resource is described by hidden variables $\lambda$ with a probability distribution $P(\lambda)$ and a set of pre-existing quantum states $\{\hat{\varsigma}_\lambda\}_\lambda$, such that the assemblage is generated by post-processing these fixed states: $\rho_{a|x}=\sum_\lambda P(\lambda)P(a|x,\lambda)\hat{\varsigma}_\lambda$, where $P(a|x,\lambda)$ encodes the reaction to Alice's outcome $a$ depending on $x$ and $\lambda$. Without loss of generality, we can write the LHS model as $\rax=\sum_\lambda P(\lambda)\delta_{a,\lambda_x}\hat{\varsigma}_\lambda$, where $\delta_{a,\lambda_x}$ is a deterministic probability distribution and $\lambda:=(\lambda_1,\lambda_2,...,\lambda_{|\mathcal{X}|})$ describes the deterministic strategy with which Alice reacts, whenever $\lambda_x=a$ for the measurement $x$~\cite{Pusey13,SNC14}. The number of all possible such deterministic strategies $\lambda$ is $|\mathcal{A}|^{|\mathcal{X}|}$. If the resource is characterized by quantum theory instead, the assemblage is obtained via $\rax=\tr_\text{A}(\Eax\otimes\openone~\rab)$. It was Shr{\"o}dinger~\cite{Schrodinger35} finding out there exist some quantum measurements $\Eax$ and quantum states $\rab$, such that the obtained assemblage does not admit a LHS model, which is called a \emph{steerable} assemblage. For brevity, we denote an assemblage like $\{\rax\}_{a,x}$ or $\{\sax\}_{a,x}$ as $\bm{\rho}$ or $\bm{\sigma}$. For the quantum state of a single system, we add a hat symbol if it is normalized: e.g., $\hatrax$ is normalized while $\rax$ is not. If $\hatrax$ are pure states for all $a,x$, then $\hatrax$ and $\{\hatrax\}_{a,x}$ can be represented by, respectively, $\ket{\hatrax}$ and $\ket{\hat{\bm{\rho}}}$, where $\hatrax=\ket{\hatrax}\bra{\hatrax}$. We denote a bipartite quantum state by $\rab$ without the hat symbol as it is always normalized in the entire manuscript.

\section{Robust self-testing of steerable assemblages}\label{Sec_robust_ST}

{\blu We need a definition of the fidelity between two quantum ensembles before giving a definiton of robust self-testing of assemblages. Consider two ensembles described by $\{p_i,\hat{\varrho}_i\}_{i=1}^N$ and $\{q_i,\hat{\varsigma_i}\}_{i=1}^N$, respectively}, where $p_i$ ($q_i$) is the probability of preparing the normalized quantum state $\hat{\varrho}_i$ ($\hat{\varsigma_i}$) with $\sum_i p_i=\sum_i q_i=1$. The fidelity between two quantum ensembles can then be defined as
\begin{equation}
F(\{p_i,\hat{\varrho}_i\},\{q_i,\hat{\varsigma_i}\})=\sum_i \sqrt{p_i q_i}~F^\text{UJ}(\hat{\varrho}_i,\hat{\varsigma}_i),
\label{Eq_F_ensembles}
\end{equation}
where $F^\text{UJ}(\hat{\varrho}_i,\hat{\varsigma}_i):=\tr\sqrt{\sqrt{\hat{\varrho}_i}\hat{\varsigma}_i\sqrt{\hat{\varrho}_i}}$ is the Uhlmann-Josza fidelity~\cite{Uhlmann76,Jozsa1994}. When $p_i=q_i$ for all $i$, the above equation recovers the typical definition of the fidelity between two ensembles (see Ref.~\cite{Liang19} and references therein). Therefore it is easy to see that $F(\{p_i,\hat{\varrho}_i\},\{q_i,\hat{\varsigma_i}\})=1$ if and only if $p_i=q_i$ and $\hat{\varrho_i}=\hat{\varsigma_i}$ for all $i$.

Now, consider a steering scenario (see Section \ref{Sec_steering}), such that for each $x$ we have two ensembles $\{\psax,\hatsax\}_a$ and $\{\prax,\hatrax\}_a$. The fidelity between them is $\sum_a\sqrt{\psax\prax}~F^\text{UJ}(\hatsax,\hatrax)$ according to Eq.~\eqref{Eq_F_ensembles}. Taking each ensemble of all the measurements $x$ into account, we can define the \emph{fidelity between two assemblages} as {\blu (see Ref.~\cite{Nery2020Distillation} for another definition of the fidelity between two assemblages)}
\begin{equation}
\begin{aligned}
&\mathcal{F}(\bm{\sigma},\bm{\rho})\\
&:=\frac{1}{|\mathcal{X}|}\sum_{a,x} \sqrt{\psax\prax}~F^\text{UJ}(\hatsax,\hatrax)\\
&=\frac{1}{|\mathcal{X}|}\sum_{a,x} \sqrt{\psax\prax}~\langle\hat{\sigma}_{a|x}|\hat{\rho}_{a|x}|\hat{\sigma}_{a|x}\rangle\\
&=\frac{1}{|\mathcal{X}|}\sum_{a,x} \sqrt{\frac{\psax}{\prax}}~\langle\hat{\sigma}_{a|x}|\rho_{a|x}|\hat{\sigma}_{a|x}\rangle.
\end{aligned}
\label{Eq_assemblage_fidelity}
\end{equation}

In a DI scheme, the underlying assemblage $\bm{\rho}$ is not characterized. Therefore, we are in general unable to compute $\mathcal{F}(\bm{\sigma},\bm{\rho})$. A strategy is to see if $\bm{\rho}$ is as useful as $\bm{\sigma}$, in the sense that all quantum information tasks using steering as resource that $\bm{\sigma}$ can accomplish is also achievable by $\bm{\rho}$. It turns out if there exists a completely positive and trace-preserving (CPTP) map $\Lambda$, such that
\begin{equation}
\Lambda(\rho_{a|x})= \sigma_{a|x}\quad\forall a,x,
\end{equation}
then $\bm{\rho}$ is as useful as $\bm{\sigma}$ in the resource theory of steering\footnote{A local CPTP map acting on the assemblage is a free operation in the resource theory of steering~\cite{Gallego15}, in the sense that any standard measure of steerability does not increase under such an action.}.

With the above, we are in a position to define the robust self-testing of assemblages:
\begin{definition} \emph{(Robust self-testing of assemblages)}

\medskip\noindent
Given an observed nonlocal correlation $\mathbf{P}\notin\mathcal{L}$ in a Bell-type experiment, we say that $\mathbf{P}$ robustly self-tests the reference assemblage $\bm{\sigma}^*$\footnote{In this paper, we use the symbol ``$*$'' behind the observables, states, or assemblages, to denote that they are the reference ones. The action of the complex conjugate is denoted by ``c.c.''.} \emph{at least} with a fidelity $f$ if for any $\bm{\rho}$ compatible with $\mathbf{P}$ there exists a CPTP map $\Lambda$, such that
\begin{equation}
\mathcal{F}(\bm{\sigma}^*,\Lambda(\bm{\rho}))\geq f,
\label{Eq_robust_ST}
\end{equation}
where $\Lambda(\bm{\rho})$ denotes $\{\Lambda(\rho_{a|x})\}_{a,x}$ for brevity.
\label{Def_1}
\end{definition}

Physically, Eq.~\eqref{Eq_robust_ST} gives a lower bound $f$ on the fidelity between the underlying assemblage and the reference assemblage in a DI setting.
The definition above is similar to a definition of self-testing of entangled states proposed in Ref.~\cite{Kaniewski16} and that of self-testing of measurements considered in Refs.~\cite{Tavakoli18b,Tavakoli2020} and \cite{Renou18} (see also Ref.~\cite{Bancal18}). In the following, we illustrate a procedure for computing $f$ for an observed nonlocal correlation $\mathbf{P}$. For the sake of simplicity, we would like to consider a typical example --- the CHSH scenario. The generalization to arbitrary scenarios can be straightforwardly obtained.

\begin{figure*}
\begin{minipage}[c]{.49\textwidth}
\includegraphics[width=5.9cm]{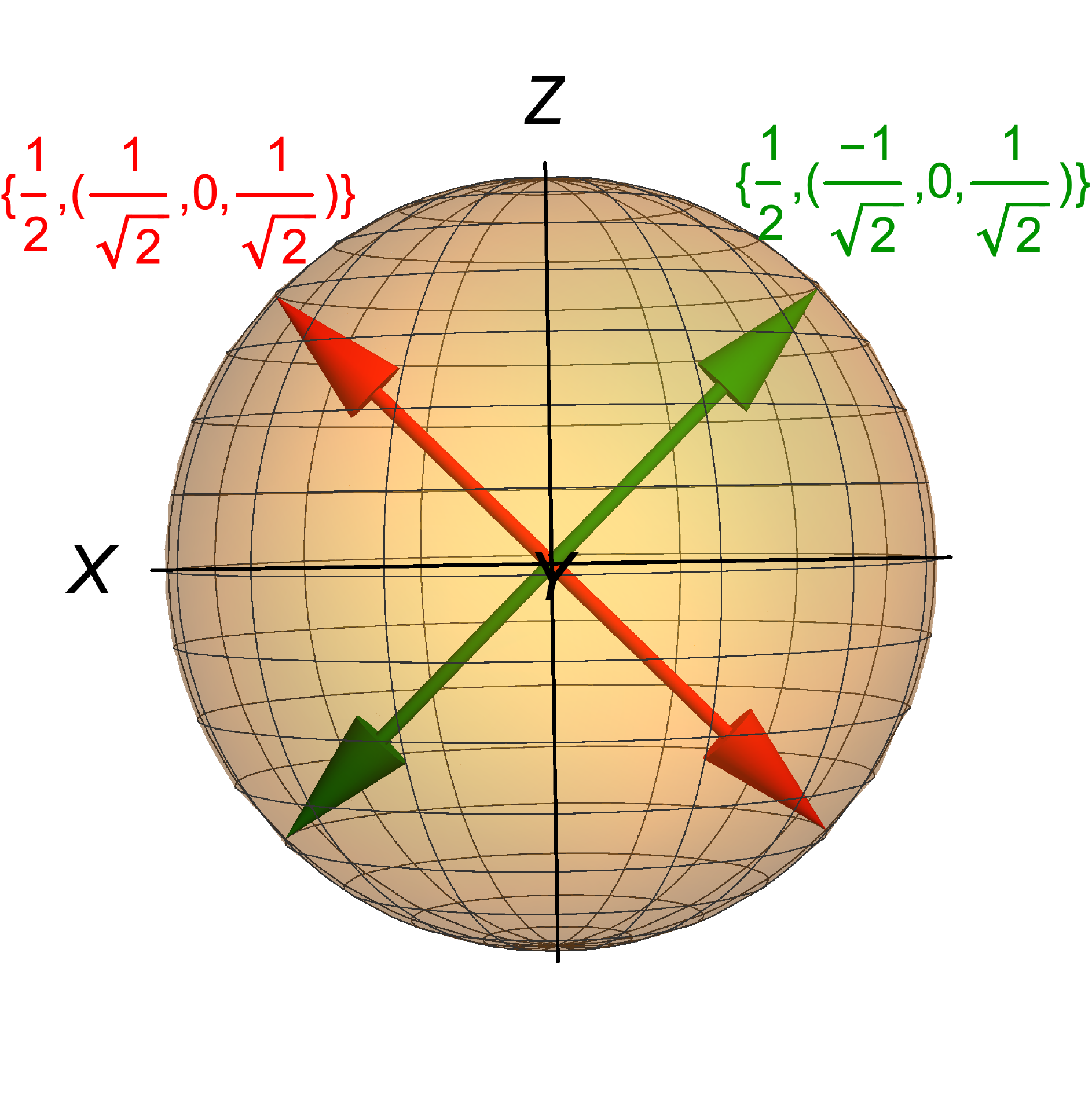}\\
\centering~\\
\text{(a)}
\end{minipage}
\begin{minipage}[c]{.49\textwidth}
\includegraphics[width=8.5cm]{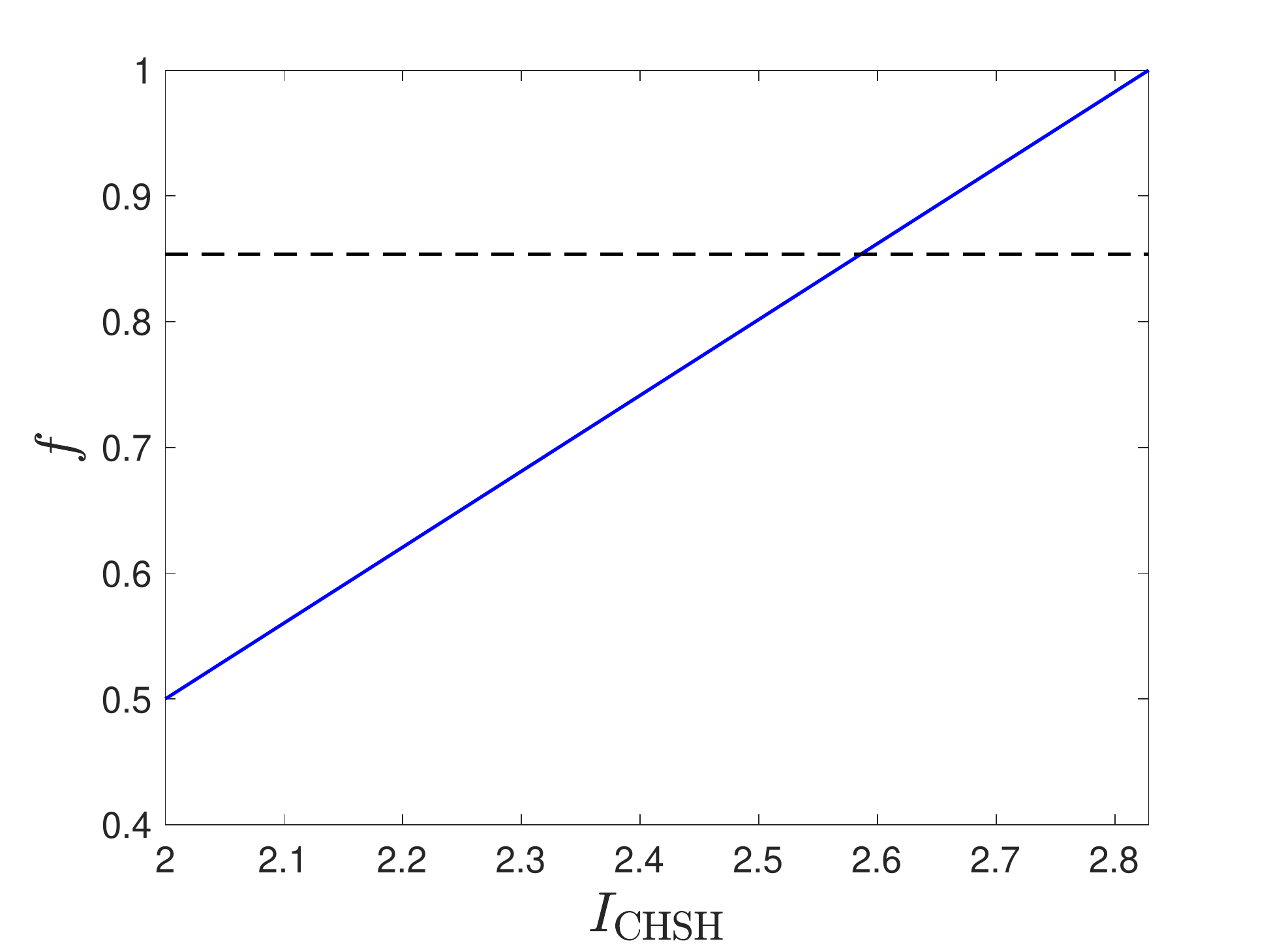}\\
\centering
\text{(b)}
\end{minipage}
\caption{
(a) The representation of the CHSH-type assemblage. Each entry of $\{\ket{\hat{\sigma}^*_{a|x}}\}_{a,x}$ described by Eq.~\eqref{Eq_CHSH_assemblage} is plotted on the $X-Z$ plane of the Bloch sphere. The red (top left - bottom right) and green (top right - bottom left) Bloch vectors represent for $\ket{\hat{\sigma}^*_{a|x}}$ with, respectively, the first ($x=1$) and second ($x=2$) measurements. For each measurement setting $x$, the array $\{(P^*_{\bm{\sigma}}(1|x),\vec{r}_{1|x}\}$ is marked on vectors corresponding to $\ket{\hat{\sigma}^*_{1|x}}$, where $\vec{r}_{1|x}$ is the Bloch vector of $\ket{\hat{\sigma}^*_{1|x}}$. (b) Robust self-testing of the CHSH-type assemblage. {\blu By solving Eq.~\eqref{Eq_min_F_SDP} with $P_{\bm{\rho}}^{\rm obs}(a|x)=P^*_{\bm{\sigma}}(a|x)$ }, we can obtain lower bounds on the fidelity between the underlying assemblage and the reference (the CHSH-type) assemblage as a function of the violation of the CHSH inequality $I_\text{CHSH}$. The horizontal dashed line with the value around $0.8536$ represents the maximal fidelity that can be achieved by a LHS model, which can be computed via Eq.~\eqref{Eq_classical_fidelity_SDP}. {\blu To carry out the computation, we use the $2$nd level of the assemblage moment matrices, where each entry is a $5$-by-$5$ matrix.}
}
\label{Fig_CHSH_type}
\end{figure*}

Consider the CHSH inequality in the correlator form~\cite{Clauser69}:
\begin{equation}
\begin{aligned}
&I_\text{CHSH}:=\\
&\langle A_1 B_1\rangle + \langle A_1 B_2\rangle + \langle A_2 B_1\rangle - \langle A_2 B_2\rangle \stackrel{\mathcal{L}}{\leq} 2,
\end{aligned}
\end{equation}
where the correlator $\langle A_x B_y\rangle:=P(a=b|x,y)-P(a\neq b|x,y)$ and its quantum realization is $\langle A_x B_y\rangle=\tr(A_x\otimes B_y \varrho^{\text{AB}})$, with $A_x$ ($B_y$) being Alice's (Bob's) observable corresponding to the $x$th ($y$th) measurement. A quantum strategy for Alice's observable $\{A_x^*\}$ and the shared state $\varrho^{*,\text{AB}}=|\psi^*\rangle^\text{AB}\langle\psi^*|$ that achieves the maximal quantum value of $2\sqrt{2}$ of the inequality is applying two Pauli observables $\{A_1^*,A_2^*\}=\{\hat{Z},\hat{X}\}$ on the maximally entangled state $|\psi^*\rangle^\text{AB}=(c/\sqrt{2})(\ket{00}-\ket{11})+(s/\sqrt{2})(\ket{01}+\ket{10})$, where $c:=\cos(\pi/8)$ and $s:=\sin(\pi/8)$.
We use this set to define the \emph{reference assemblage} by using the relation $\sax=\tr_{\text{A}}( E_{a|x}\otimes\openone~\rab)$, hence obtain each entry of the reference assemblage denoted by $\sigma_{a|x}^*=\psaxstar\ket{\hat{\sigma}^*_{a|x}}\bra{\hat{\sigma}^*_{a|x}}$ with $\psaxstar=1/2$ for all $a,x$ and
\begin{equation}
\begin{aligned}
&|\hat{\sigma}_{1|1}^*\rangle=c|0\rangle+s|1\rangle,~
|\hat{\sigma}_{1|2}^*\rangle=-c|0\rangle+s|1\rangle,\\
&|\hat{\sigma}_{2|1}^*\rangle=s|0\rangle-c|1\rangle,~
|\hat{\sigma}_{2|2}^*\rangle=s|0\rangle+c|1\rangle
\end{aligned}
\label{Eq_CHSH_assemblage}
\end{equation}
being four normalized quantum states steered on Bob's side. We refer to such an assemblage as ``the CHSH-type assemblage''\footnote{We don't use the term ``the maximally steerable assemblage'' due to the non-existence of \emph{steering bit}~\cite{Gallego15}. In other words, there is no measure-independent maximally steerable assemblage.}. As seen in Fig.~\ref{Fig_CHSH_type}(a), in this work we provide a visual representation of each type of qubit assemblage that we would like to self-test. Note the above maximally entangled state $|\psi^*\rangle^\text{AB}$ is unitarily equivalent to the singlet. {\blu By using the form of $|\psi^*\rangle^\text{AB}$,} Bob's optimal (not necessarily be unique) quantum strategy is the same as Alice's observables, i.e., $B_1^*=A_1^*=\hat{Z}$ and $B_2^*=A_2^*=\hat{X}$. The next step is to choose a proper CPTP map $\Lambda$ in Definition~\ref{Def_1}. It is rather convenient for us to use the Choi-Jamio{\l}kowski (CJ) isomorphism~\cite{Jamiokowski74,Choi75} so that the task of finding a CPTP map can be transformed to finding a positive-semidefinite matrix $\Omega\succeq 0$, where $\Omega=(\openone\otimes\Lambda)|\phi^+\rangle\langle\phi^+|$ is the so-called \emph{CJ matrix} with $|\phi^+\rangle=\sum_i|i\rangle_\text{B}\otimes|i\rangle_{\text{B}'}$ being the unnormalized maximally entangled state. More specifically, we have the following relation
\begin{equation}
\Lambda(\rho_{a|x}) = \tr_{\text{B}}\big[\Omega(\rho_{a|x}^{\mathsf{T}}\otimes\openone)\big]\quad\forall a,x,
\end{equation}
where $\mathsf{T}$ denotes the action of transpose with respect to the computational basis. For later use, let us apply the transpose on the above equation:  $\big[\Lambda(\rho_{a|x})\big]^\mathsf{T} = \tr_{\text{B}}\big[(\rho_{a|x}\otimes\openone)\Omega^\mathsf{T} \big]$. When performing the optimal quantum strategy, namely $A_x^*$ and $|\psi^*\rangle^\text{AB}$, we have $\bm{\rho}=\bm{\sigma}^*$. Thus we can choose the following CJ matrix: $\Omega^\mathsf{T} = |\phi^+\rangle\langle\phi^+|$, i.e., the corresponding CPTP map $\Lambda$ is the identity map. Then, $\Omega^\mathsf{T}$ can be represented by Bob's optimal observables $\{B_y^*\}$ as following\footnote{\blu Similar with finding a representation of the swap gate in the swap method~\cite{Yang14,Bancal15}, here we also have infinite ways of the representation of the CJ matrix.}}:
\begin{equation}
\begin{aligned}
\Omega^\mathsf{T}&=\frac{\openone+B_1^*}{2}\otimes\ket{0}\bra{0}+\frac{B_2^*-B_2^*B_1^*}{2}\otimes\ket{0}\bra{1}\\
&+\frac{B_2^*-B_1^*B_2^*}{2}\otimes\ket{1}\bra{0}+\frac{\openone-B_1^*}{2}\otimes\ket{1}\bra{1}.
\end{aligned}
\label{Eq_Omega}
\end{equation}
In a DI setting, Bob's observables are not characterized, therefore we relax $B_1^*$ and $B_2^*$ to unknown unitary and Hermitian observables $B_1$ and $B_2$. Finally, we obtain a DI description of the fidelity:
\begin{equation}
\begin{aligned}
&\mathcal{F}\big(\bm{\sigma^*},\Lambda(\bm{\rho})\big)=\mathcal{F}\Big((\bm{\sigma}^*)^\mathsf{T},\big[\Lambda(\bm{\rho})\big]^\mathsf{T}\Big)\\
&= \frac{1}{2}\sum_{a,x}\sqrt{\frac{\psaxstar}{\prax}}~\bra{\hat{\sigma}_{a|x}^{*,{\rm c.c.}}} \big[\Lambda\big({\rho_{a|x}}\big)\big]^\mathsf{T} \ket{\hat{\sigma}_{a|x}^{*,{\rm c.c.}}}\\
&= \frac{1}{2}\sum_{a,x}\sqrt{\frac{\psaxstar}{\prax}}~\bra{\hat{\sigma}_{a|x}^*} \tr_{\text{B}}\big[(\rho_{a|x}\otimes\openone)\Omega^{\mathsf{T}}\big] \ket{\hat{\sigma}_{a|x}^*},
\end{aligned}
\label{Eq_DI_fidelity_CHSH}
\end{equation}
where the first equality holds since the fidelity between a pure and a mixed state is equivalent to that between their transpose. The notation ${\rm c.c.}$ in the second line denotes the complex conjugate, and $\ket{\hat{\sigma}_{a|x}^{*,{\rm c.c.}}}=\ket{\hat{\sigma}_{a|x}^{*}}$ since they are all real in the CHSH case. Note that due to the trace-preserving property of $\Lambda$, we have $\prax:=\tr(\rax)=\tr(\Lambda(\rax))$.

Having a DI fidelity $\mathcal{F}\big(\bm{\sigma^*},\Lambda(\bm{\rho})\big)$, our goal is to find its lower bound described in Definition~\ref{Def_1} for any $\bm{\rho}$ compatible with the observed quantum violation $I_\text{CHSH}^\text{obs}$. In other words, the problem is formulated as:
\begin{equation}
\begin{aligned}
\min_{\mathbf{P}}\quad &\mathcal{F}\big(\bm{\sigma^*},\Lambda(\bm{\rho})\big),\\
\text{such that}\quad & I_\text{CHSH}(\mathbf{P}) = I_\text{CHSH}^\text{obs},\\
& \mathbf{P}\in\mathcal{Q},
\end{aligned}
\label{Eq_min_F}
\end{equation}
where $\mathcal{Q}$ is the quantum set of correlations mentioned in Section~\ref{Sec_Bell}. Since there is no known simple way to characterize $\mathbf{P}\in\mathcal{Q}$, a compromise strategy is to use the outer approximation~\cite{NPA,NPA2008,Doherty08} to characterize a superset of $\mathcal{Q}$. Some variations~\cite{Moroder13,CBLC16,CMBC2021} were also proposed to tackle specific DI problems, and it is rather convenient to use the so-called the \emph{assemblage moment matrices}~\cite{CBLC16,CBLC18} to approximate the quantum set $\mathcal{Q}$. The reason is that if we look closely at the objective function $\mathcal{F}\big(\bm{\sigma^*},\Lambda(\bm{\rho})\big)$ (which is obtained by substituting $\Omega^\mathsf{T}$ in Eq.~\eqref{Eq_DI_fidelity_CHSH} with Eq.~\eqref{Eq_Omega}), it is a polynomial where each term is of forms such as $\tr(\rax)$, $\tr(B_1\rax)$, $\tr(B_1B_2\rax)$, etc, up to some coefficients. In general, they can be described as $\tr(S_i^\dag S_j \rho_{a|x})$, which are exactly entries of the assemblage moment matrices, with $\{S_i\}$ being the union of the identity, Bob's unknown observables, and their products, i.e., $\{S_i\}=\{\openone, B_1, B_2, B_1B_2,...\}$. The CHSH inequality violation $I_\text{CHSH}$ can also be expressed as a linear combination of this form, i.e., $\langle A_x B_y\rangle=\tr\big[B_y(\rho_{1|x}-\rho_{2|x})\big]$. Therefore, the constraint $\mathbf{P}\in\mathcal{Q}$ in Eq.~\eqref{Eq_min_F} is relaxed and the problem now becomes
\begin{equation}
\begin{aligned}
\min \quad &\mathcal{F}\big(\bm{\sigma^*},\Lambda(\bm{\rho})\big)\\
\text{s.t.} \quad &I_\text{CHSH}(\mathbf{P}) = I_\text{CHSH}^\text{obs}\\
& {\blu P_{\bm{\rho}}(a|x)=P_{\bm{\rho}}^{\rm obs}(a|x) }, \\
& \chi[\rho_{a|x},\mathcal{S}]\succeq 0\quad\forall a,x,\\
& \sum_a\chi[\rho_{a|x},\mathcal{S}] = \sum_a\chi[\rho_{a|x'},\mathcal{S}]\quad\forall x\neq x',\\
\end{aligned}
\label{Eq_min_F_SDP}
\end{equation}
where $\chi[\rho_{a|x},\mathcal{S}]:=\sum_{i,j}\ket{i}\bra{j}\tr(S_j^\dag S_i \rho_{a|x})$ are the assemblage moment matrices~\cite{CBLC16,CBLC18}, and $\mathcal{S}:=\{S_i\}$ is a sequence of Bob's unknown observables and their products. {\blu {\blub The second constraint makes the DI expression of the fidelity a linear function of the free variables, so that the problem is a semidefinite program (SDP)~\cite{BoydBook}}, which can be solved efficiently with computer packages. Note that $P^{\rm obs}_{\bm{\rho}}(a|x)=P_{\bm{\sigma}}^*(a|x)$ is the only choice~\cite{Goh18} for the maximal quantum violation of the CHSH inequality and the tilted CHSH inequality in the next section.} In the above SDP, the minimization is taken over the free variables of the moment matrices, e.g., $\tr(B_2B_1\rax)$, $\tr(B_2B_1B_1\rax)$, etc. The {\blub third and fourth} constraints characterize a superset of the quantum set $\mathcal{Q}$~\cite{CBLC16,CBLC18}, therefore the solution, denoted by $f$, is a lower bound on {\blub the solution} of Eq.~\eqref{Eq_min_F}. 

In Fig.~\ref{Fig_CHSH_type}(b), we plot $f$ as a function of the observed quantum violations of the CHSH inequality $I_\text{CHSH}^\text{obs}$. As can be seen, the fidelity achieves the value of $1$ with the maximal quantum violation of $2\sqrt{2}$. This means we successfully self-test that the underlying assemblage contains the CHSH-type one. To obtain the ``classical fidelity'' represented by the horizontal line in the figure, we consider that Bob's task is to simulate the reference assemblage without any steerable resource. Indeed, he is able to discard his share and prepare a fixed state from the set $\{\hat{\varsigma_\lambda}\}$, which turns out the assemblage produced will admit a LHS model described in Sec.~\ref{Sec_steering}. Hence, the maximal overlap with the reference assemblage that a LHS model can achieve is $\max\{  \mathcal{F}\big(\bm{\sigma^*},\bm{\rho}\big) | \rho_{a|x}=\sum_\lambda P(\lambda)\delta_{a,\lambda_x}\hat{\varsigma}_\lambda\}$. It can be computed via the following SDP (see Appendix \ref{SecApp_classical_fidelity} for the derivation):
\begin{equation}
\begin{aligned}
\max_{\{\hat{\varsigma}_\lambda\}} \quad & \frac{\sqrt{|\mathcal{A}|}}{|\mathcal{X}||\lambda|}\sum_{a,x,\lambda}\sqrt{\psaxstar} \delta_{a,\lambda_x}\tr(\hatsaxstar\hat{\varsigma})\\
\text{s.t.}\quad & \hat{\varsigma}_\lambda\succeq 0,\quad \tr(\hat{\varsigma}_\lambda)=1 \quad\forall\lambda.
\end{aligned}
\label{Eq_classical_fidelity_SDP}
\end{equation}
where $|\lambda|=|\mathcal{A}|^{|\mathcal{X}|}$ is the number of elements of all the vectors $\lambda$.

{\blu Having provided the self-testing statement for the CHSH-type assemblage, we summarize the general procedure of self-testing assemblages {\blub in the following grey box}: }
\begin{tcolorbox}
\begin{itemize}
\item Step 1. Assign the reference assemblage (cf., Eq.~\eqref{Eq_CHSH_assemblage}).
\item Step 2. Choose an expression of the CJ matrix associated with the identity map (cf., Eq.~\eqref{Eq_Omega}).
\item Step 3. Relax the scheme to a DI scenario (cf., relax $B_y^*$ to $B_y$).
\item Step 4. Obtain a DI description of the fidelity (cf. Eq.~\eqref{Eq_DI_fidelity_CHSH}).
\item Step 5. Compute a lower bound on the fidelity via SDP (cf. Eq.~\eqref{Eq_min_F_SDP}).
\item Step 6. Compute the trivial fidelity via SDP (cf. Eq.~\eqref{Eq_classical_fidelity_SDP}).
\end{itemize}
\end{tcolorbox}

We would like to point out that the procedure of the relaxation is somehow similar to the one in the swap method~\cite{Yang14,Bancal15}. While the swap method relaxes the characterization of the unitary circuit to self-test the reference entangled state, here we relax the characterization of the CJ matrix to self-test the reference steerable assemblage. As can be seen in Sec.~\ref{Sec_EB_type}, our method further enables one to self-test assemblages containing complex entries.

\begin{figure*}
\begin{minipage}[c]{.49\textwidth}
\includegraphics[width=5.9cm]{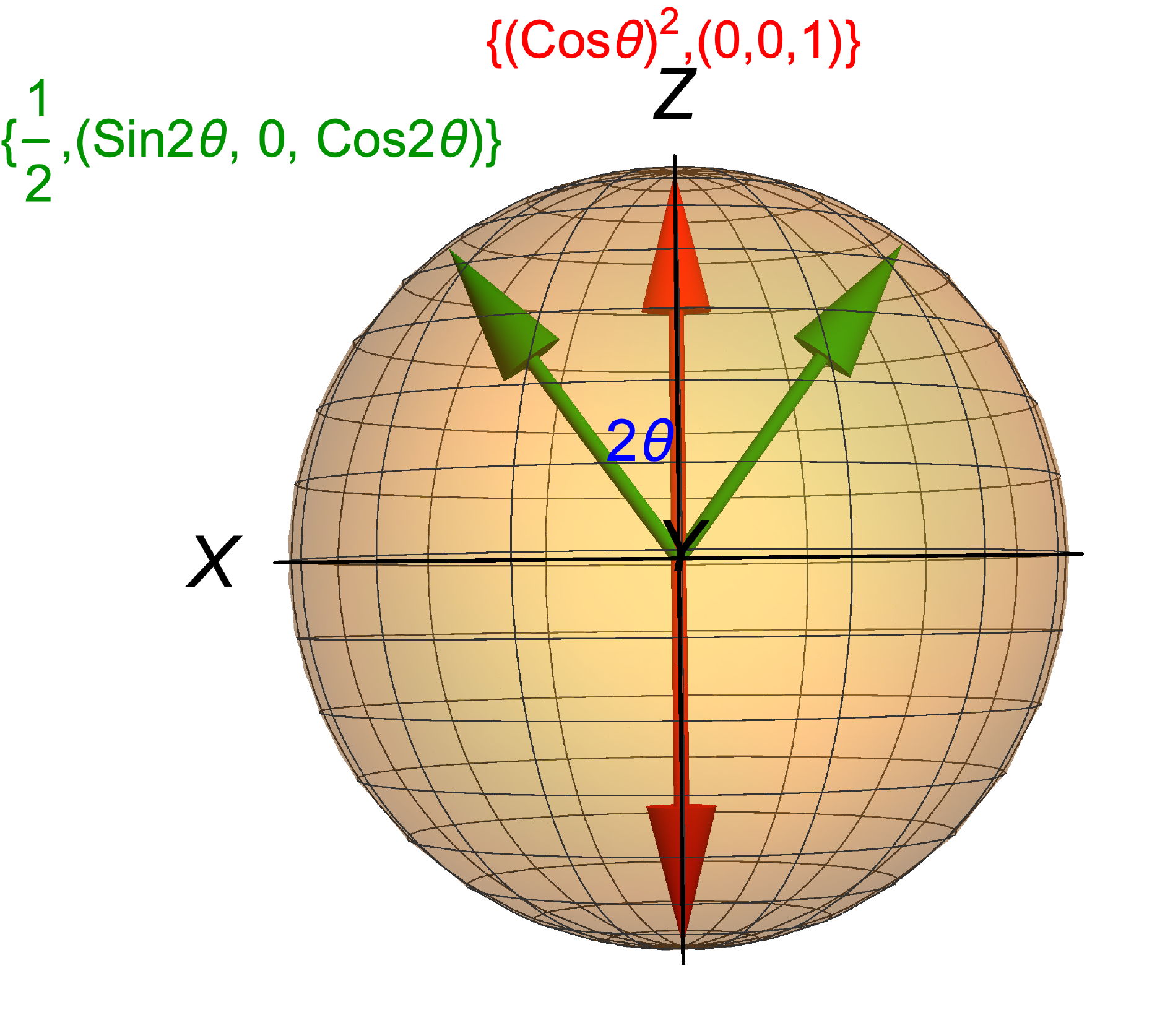}\\
\centering
\text{(a)}\\
\includegraphics[width=8cm]{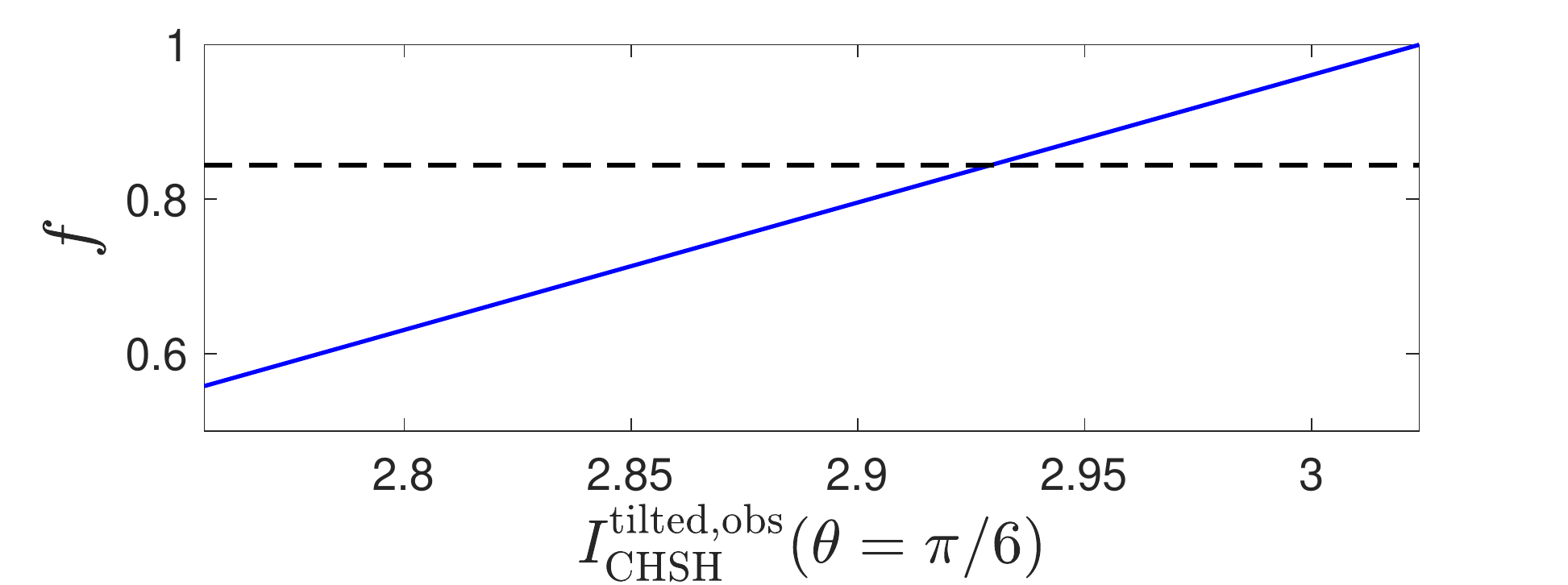}\\
\centering
\text{(b)}
\end{minipage}
\begin{minipage}[c]{.49\textwidth}
~\\~\\~\\
\includegraphics[width=8cm]{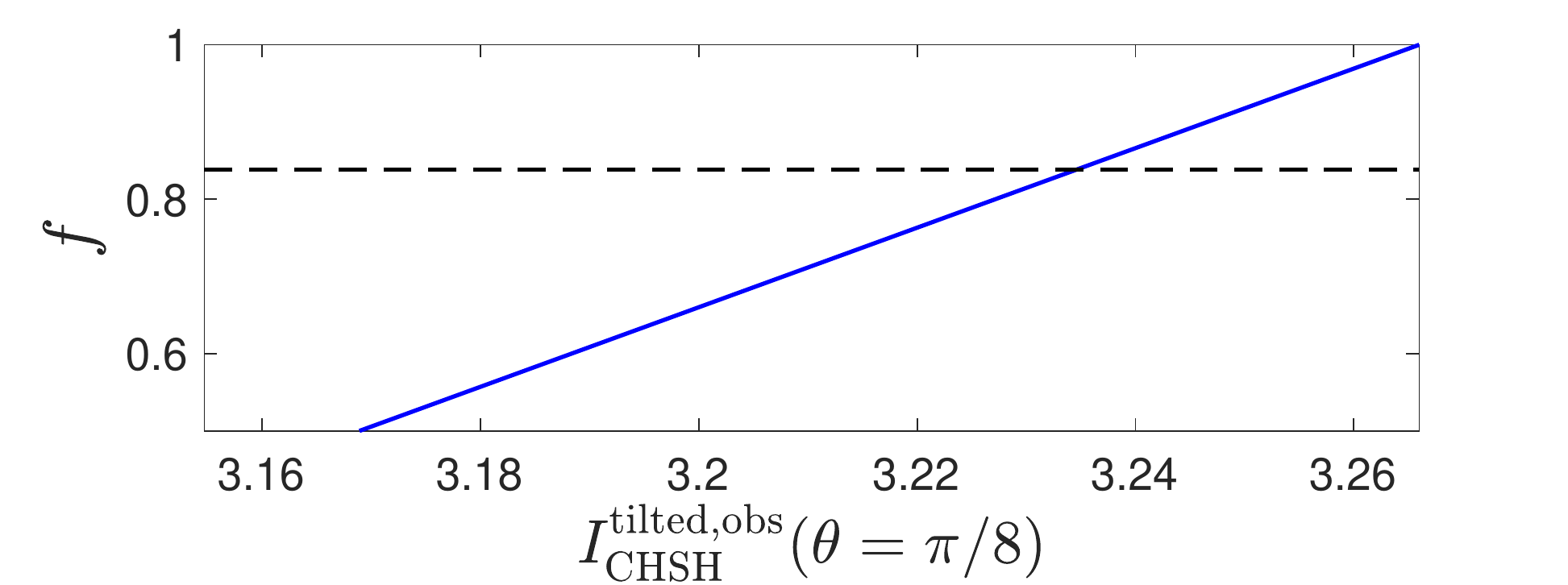}\\
\centering
\text{(c)}\\~\\~\\~\\
\includegraphics[width=8cm]{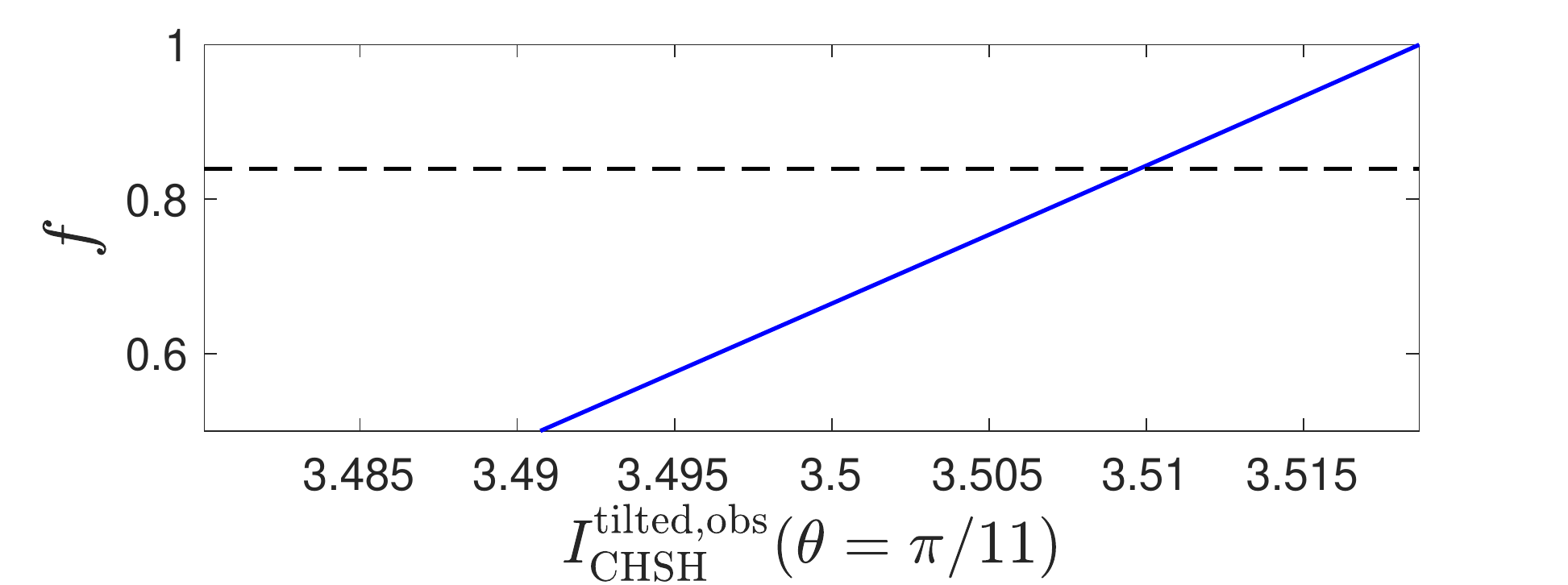}\\
\centering
\text{(d)}\\
\end{minipage}
\caption{
(a) The representation of the tilted-CHSH-type assemblage. The angle between the Bloch vectors of $\ket{\hat{\sigma}_{1|1}^*}$ and $\ket{\hat{\sigma}_{1|2}^*}$, marked as $2\theta$, is the parameter of the shared state $\ket{\psi^*}^{\text{AB}}=\cos\theta\ket{00}+\sin\theta\ket{11}$. (b-d) {\blu By solving Eq.~\eqref{Eq_min_F_SDP_tilted_CHSH} with $P^{\rm obs}_{\bm{\rho}}(a|x)=P_{\bm{\sigma}}^*(a|x)$, we can obtain DI lower bounds on the fidelity between the underlying and the reference assemblage as a function of the tilted CHSH inequality $I_{\rm CHSH}^{\rm tilted,obs}$}. The horizontal-dashed lines represent the classical fidelities, which are respectively around $0.8440$, $0.8385$, and $0.8397$. Each $x$-axis of the plots ranges from the local bound to the quantum bound of $I_{\rm CHSH}^{\rm tilted}$. {\blu To carry out the computation, we use assemblage moment matrices whose sequence is with size of $29$ (see Appendix~\ref{SecApp_Sequence} for the detail).}
}
\label{Fig_tilted_CHSH_type}
\end{figure*}

\section{Robust self-testing of other types of assemblages}
In this section, we consider other paradigmatic Bell scenarios and obtain the robust self-testing statement in each of them. In particular, we raise these scenarios for specific purpose: Section~\ref{Sec_tilted_CHSH_type} stands for the situation where Alice's and Bob's measurements achieving the maximal quantum violation of the given Bell inequality are chosen to be different, hence we have to use the tool of \emph{localizing matrices}~\cite{Pironio10b}. In Sec.~\ref{Sec_EB_type}, {\blu we show that our method can be used for self-testing complex-valued assemblages.}

\subsection{The tilted CHSH type}\label{Sec_tilted_CHSH_type}
The tilted CHSH inequality is written as~\cite{Acin12} (see also \cite{Yang13,Bamps15})
\begin{equation}
\begin{aligned}
&I_{\rm CHSH}^{\rm tilted}:=\\
&\alpha\langle A_1\rangle  + \langle A_1 B_1\rangle + \langle A_1 B_2\rangle + \langle A_2 B_1\rangle - \langle A_2 B_2\rangle\\
&\stackrel{\mathcal{L}}{\leq} 2+\alpha,
\end{aligned}
\end{equation}
where $\alpha\in[0,2)$. The maximal quantum violation of the inequality is given by $\sqrt{8+2\alpha^2}$, and can be achieved by Alice performing the measurements $\{A_1^*,A_2^*\}=\{\hat{Z},\hat{X}\}$ on the shared state $\ket{\psi^*}^\text{AB} = \cos\theta\ket{00}+\sin\theta\ket{11}$, with $\sin 2\theta=\sqrt{\frac{4-\alpha^2}{4+\alpha^2}}$. With this quantum strategy, we then define Bob's reference assemblage, referred to the \emph{tilted-CHSH-type} assemblage, as $\sigma_{a|x}^*=\psaxstar\ket{\hat{\sigma}^*_{a|x}}\bra{\hat{\sigma}^*_{a|x}}$ with
\begin{equation}
\begin{aligned}
&P_{\bm{\sigma}}^*(1|1) = \cos^2\theta,\quad P_{\bm{\sigma}}^*(1|2)=\frac{1}{2}\\
&P_{\bm{\sigma}}^*(2|1) = \sin^2\theta,\quad P_{\bm{\sigma}}^*(2|2)=\frac{1}{2},
\end{aligned}
\end{equation}
and
\begin{equation}
\begin{aligned}
&|\hat{\sigma}_{1|1}^*\rangle=|0\rangle,\quad
|\hat{\sigma}_{1|2}^*\rangle=\cos\theta|0\rangle+\sin\theta|1\rangle,\\
&|\hat{\sigma}_{2|1}^*\rangle=|1\rangle,\quad
|\hat{\sigma}_{2|2}^*\rangle=\cos\theta|0\rangle-\sin\theta|1\rangle,
\end{aligned}
\label{Eq_tilted_CHSH_assemblage}
\end{equation}
the representation of which is plotted in Fig.~\ref{Fig_tilted_CHSH_type}(a). To achieve the quantum bound of $I_{\rm CHSH}^{\rm tilted}$, Bob's observables can be chosen as $B_1^*=\cos\mu\hat{Z} + \sin\mu\hat{X}$ and $B_2^*=\cos\mu\hat{Z} - \sin\mu\hat{X}$, with $\mu=\arctan(\sin 2\theta)$. There is a tricky point we have to take care of when performing the relaxation to a DI scenario: If we relax, say, $\hat{Z}=(B_1^*+B_2^*)/\cos\mu$ to an unknown observable $(B_1+B_2)/\cos\mu$, then it is in general not a unitary. To tackle this problem, we use the same technique as in Refs.~\cite{Yang14,Bancal15}. For any operator $B$, there exists a unitary operator $U$ such that $UB$ is positive semidefinite, i.e., this is simply the polar decomposition. Moreover, $U=B^\dag$ if $B$ is unitary. Therefore, we introduce observables $B_3^*$ and $B_4^*$, such that
\begin{equation}
\begin{aligned}
&B_3^*\left(\frac{B_1^*+B_2^*}{\cos\mu}\right)\succeq 0,\quad
B_4^*\left(\frac{B_1^*-B_2^*}{\sin\mu}\right)\succeq 0.
\end{aligned}
\label{Eq_polar_decomposition_tilted_CHSH}
\end{equation}
Since the entire term in each pair of parentheses is unitary, we have $B_3^*=\hat{Z}$ and $B_4^*=\hat{X}$.
By this, we can represent the CJ matrix with these observables:
\begin{equation}
\begin{aligned}
\Omega^\mathsf{T}&=\frac{\openone+B_3^*}{2}\otimes\ket{0}\bra{0}+\frac{B_4^*-B_4^*B_3^*}{2}\otimes\ket{0}\bra{1}\\
&+\frac{B_4^*-B_3^*B_4^*}{2}\otimes\ket{1}\bra{0}+\frac{\openone-B_3^*}{2}\otimes\ket{1}\bra{1}.
\end{aligned}
\label{Eq_opt_Omega_tilted_CHSH}
\end{equation}
In a DI setting, $B_y^*$ are relaxed to uncharacterized unitary and Hermitian observables $B_y$. Consequently, we obtain a DI description of the fidelity $\mathcal{F}\big(\bm{\sigma}^*,\Lambda(\bm{\sigma})\big)$ between the underlying assemblage $\bm{\sigma}$ and the reference one $\bm{\sigma}^*$ (c.f., Eq.~\eqref{Eq_DI_fidelity_CHSH}). As in the CHSH case, given a quantum violation of $I_{\rm CHSH}^{\rm tilted}$, a lower bound on the fidelity can also be computed by the following SDP:
\begin{equation}
\begin{aligned}
\min \quad &\mathcal{F}\big(\bm{\sigma^*},\Lambda(\bm{\sigma})\big)\\
\text{s.t.} \quad &I_{\rm CHSH}^{\rm tilted}(\mathbf{P}) = I_{\rm CHSH}^{\rm tilted,obs}\\
& {\blu P_{\bm{\rho}}(a|x)=P_{\bm{\rho}}^{\rm obs}(a|x) },\\
& \chi[\sax,\mathcal{S}]\succeq 0,\\
& \sum_a\chi[\sax,\mathcal{S}] = \sum_a\chi[\sigma_{a|x'},\mathcal{S}],\\
&\chi_\text{\tiny L}\big[\sax,\frac{B_3(B_1+B_2)}{\cos\mu},\mathcal{S}'\big]\succeq 0,\\
&\chi_\text{\tiny L}\big[\sax,\frac{B_4(B_1-B_2)}{\sin\mu},\mathcal{S}'\big]\succeq 0.
\end{aligned}
\label{Eq_min_F_SDP_tilted_CHSH}
\end{equation}
Compared with Eq.~\eqref{Eq_min_F_SDP}, two more constraints (the last two lines) are included, which are relaxations of Eq.~\eqref{Eq_polar_decomposition_tilted_CHSH}. The term $\chi_\text{\tiny L}[\sax,B,\mathcal{S}']:=\sum_{ij}\ket{i}\bra{j}\tr[(S'_j)^\dag B S'_i\sax]$ is a variant of the so-called \emph{localizing matrix}~\cite{Pironio10b,Yang14,Bancal15}, and, by construction, it is positive semidefinite if $B\succeq 0$\footnote{More specifically, we can think of the localizing matrix as a completely positive map on $\sax$, namely $\mathcal{E}(\sax)=\sum_n K_n\sax K_n^\dag$ with $K_n:=\sum_i\ket{i}\bra{n}B^{1/2}S_i$, where $(B^{1/2})^\dag B^{1/2}=B$ is a Cholesky decomposition of $B$ (see Ref.~\cite{Moroder13} for treating the standard moment matrix as a completely positive map.)}. The requirement of the second sequence, $\mathcal{S}'$, is that the localizing matrices $\chi_\text{\tiny L}$ contain all the moment terms of the DI fidelity but cannot contain {\blub terms} not shown in $\chi$, i.e., $S' \subseteq S$. The results are plotted in Figs.~\ref{Fig_tilted_CHSH_type}(b-d).

\begin{figure*}
\begin{minipage}[c]{.49\textwidth}
\includegraphics[width=7cm]{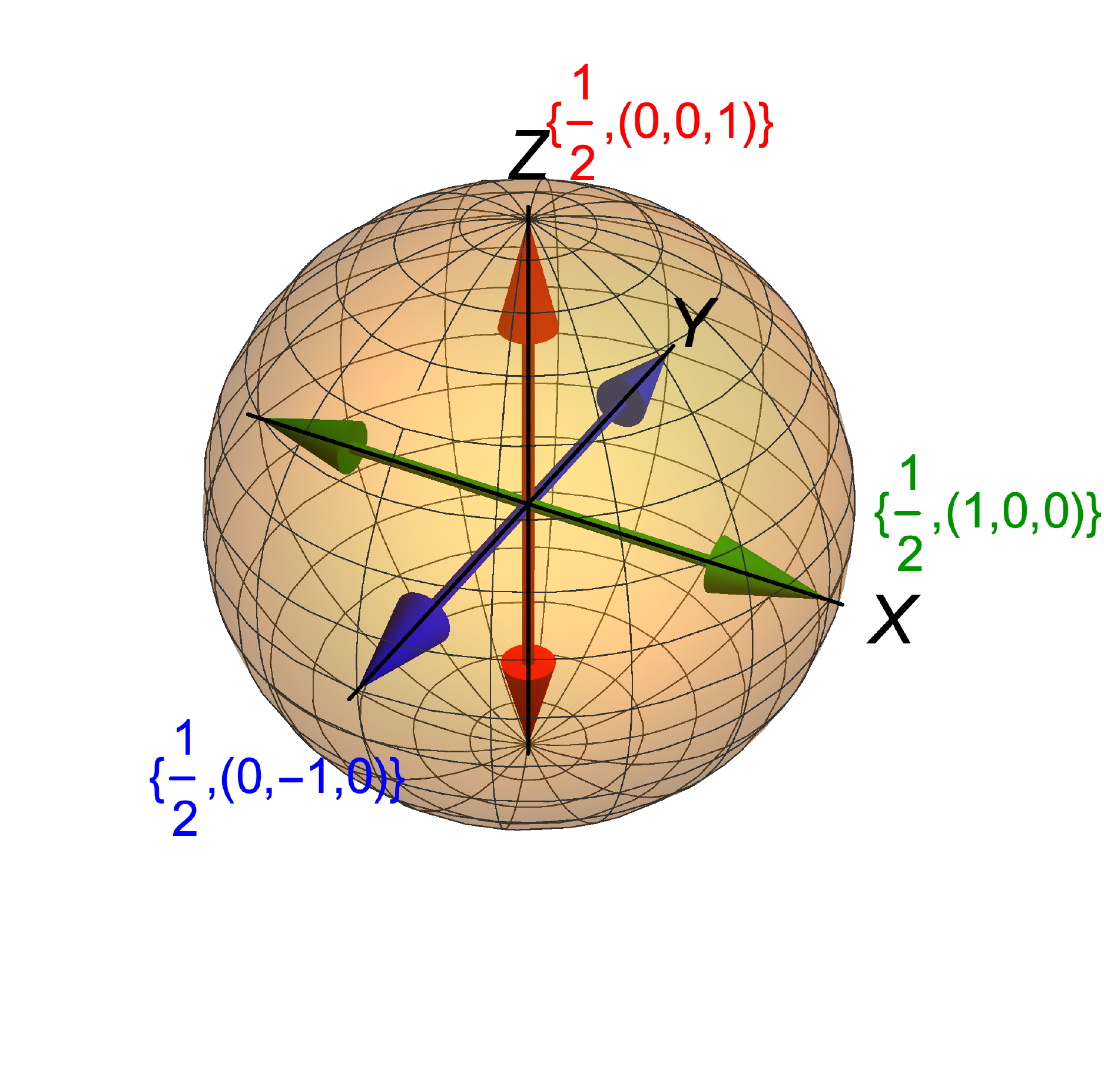}\\
\centering
\text{(a)}
\end{minipage}
\begin{minipage}[c]{.49\textwidth}
\includegraphics[width=8.5cm]{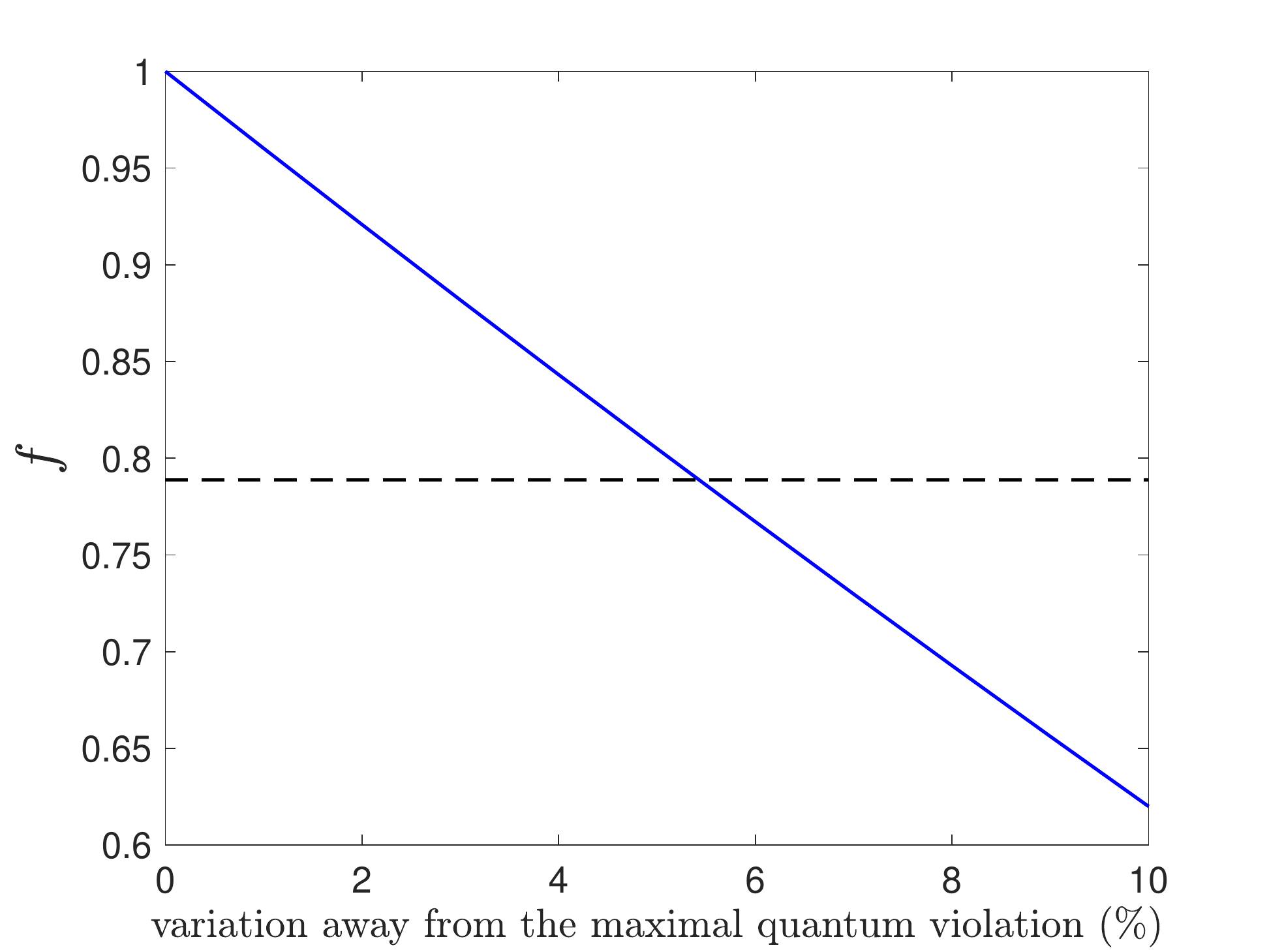}\\
\quad\\
\centering
\text{(b)}
\end{minipage}
\caption{
(a) The representation of the elegant-Bell-type (and the $I_{3622}$-type) assemblage described in Eq.~\eqref{Eq_EB_assemblage}. {\blubb To self-test a complex-valued assemblage, one should compute the fidelity between the underlying assemblage and the controlled mixture of the reference assemblage and its transpose (cf. Def.~\ref{Def_2}).} (b) {\blu By solving Eq.~\eqref{Eq_min_F_SDP_EBI}, we can obtain DI lower bounds on the fidelity as a function of the percentage variation away from the maximal quantum violation of the elegant Bell inequality.
The horizontal dashed line with the value around $0.7887$ represents the maximal fidelity that can be achieved by a LHS model, which can be computed via Eq.~\eqref{Eq_classical_fidelity_SDP}. To carry out the computation, we choose $P_{\bm{\rho}}^{\rm obs}(a|x)=P^*_{\bm{\sigma}}(a|x)$} and chose  sequences with the length of $38$ (see Appendix~\ref{SecApp_Sequence} for the detail). {\blubb The result for self-testing the same assemblage with $I_{3622}$ inequality is attached in Appendix \ref{SecApp_EBIandI3622}.}
}
\label{Fig_EB_and_I3622}
\end{figure*}

\subsection{{\blu Self-testing of complex-valued assemblages:} the elegant-Bell type and the $I_{3622}$ type}\label{Sec_EB_type}
{\blubb In this section, we self-test the reference assemblage by using the elegant Bell inequality~\cite{Gisin:ManyQuestions} and the $I_{3622}$ inequality~\cite{Acin16}. In particular, since the reference assemblage contains complex-valued entries, we have to properly revise the definition of self-testing of assemblages in Definition~\ref{Def_1}. The reason is the same as the situation in self-testing of measurements. That is, one cannot distinguish a complex assemblage $\{\sigma_{a|x}\}$ from its transpose $\{\sigma_{a|x}^\mathsf{T}\}$ in a DI setting due to the invariance of the correlation under the transpose:
\begin{equation}
P(a,b|x,y) = \tr(E_{b|y} \sigma_{a|x}) =  \tr(E_{b|y}^\mathsf{T} \sigma_{a|x}^\mathsf{T}).
\label{Eq_invariance}
\end{equation}
More importantly, there is no CPTP map which maps a complex-valued assemblage $\{\sigma_{a|x}\}$ to its transpose $\{\sigma_{a|x}^\mathsf{T}\}$. In fact, any controlled mixture of $\{\sigma_{a|x}\}$ and $\{\sigma_{a|x}^\mathsf{T}\}$ will lead to the same correlation (by performing the controlled mixture of $E_{b|y}$ and $E_{b|y}^\mathsf{T}$ on them). Therefore, if we treat the family of controlled-mixture assemblages as equivalent in a DI scheme, we should generalize the definition of the usefulness of the assemblage: we say that the underlying assemblage $\bm{\rho}$ is as useful as the complex-valued reference assemblage $\bm{\sigma}^*$ if there exists a CPTP map $\Lambda$, a real number $0\leq q\leq 1$, and an orthornormal set $\{\ket{0},\ket{1}\}$, such that
\begin{equation}
\Lambda(\rho_{a|x})=q\sigma_{a|x}^*\otimes|0\rangle\langle 0| + (1-q) (\sigma_{a|x}^*)^\mathsf{T}\otimes|1\rangle\langle 1|,
\end{equation}
for all $a,x$. With this, we arrive the following definition of robust self-testing of complex-valued assemblages (cf. Definition.~\ref{Def_1}).

\begin{definition} \emph{(Robust self-testing of complex-valued assemblages)}

\medskip\noindent
Given an observed nonlocal correlation $\mathbf{P}\notin\mathcal{L}$ in a Bell-type experiment, we say that $\mathbf{P}$ robustly self-tests the complex-valued reference assemblage $\bm{\sigma}^*$ \emph{at least} with a fidelity $f$ if for any $\bm{\rho}$ compatible with $\mathbf{P}$ there exists a CPTP map $\Lambda$, a real number $0\leq q\leq 1$, and an orthornormal set $\{\ket{0},\ket{1}\}$, such that
\begin{equation}
\begin{aligned}
&\mathcal{F}\big[q\sigma_{a|x}^*\otimes|0\rangle\langle 0| + (1-q) (\sigma_{a|x}^*)^\mathsf{T}\otimes|1\rangle\langle 1|,\Lambda(\rho_{a|x})\big]\\
&\geq f\quad \forall a,x.
\end{aligned}
\label{Eq_robust_ST_complex}
\end{equation}
\label{Def_2}
\end{definition}

}

{\blubb In what follows, we demonstrate how to self-test elegant-Bell-type assemblage with Definition \ref{Def_2}.}
The elegant Bell inequality is written as~\cite{Gisin:ManyQuestions} (see also Ref.~\cite{Christensen15}):
\begin{equation}
\begin{aligned}
I_{\rm E}&:= \langle A_1 B_1\rangle + \langle A_1 B_2\rangle - \langle A_1 B_3\rangle - \langle A_1 B_4\rangle\\
&+\langle A_2 B_1\rangle - \langle A_2 B_2\rangle + \langle A_2 B_3\rangle - \langle A_2 B_4\rangle\\
&+\langle A_3 B_1\rangle - \langle A_3 B_2\rangle - \langle A_3 B_3\rangle + \langle A_3 B_4\rangle \stackrel{\mathcal{L}}{\leq} 6.
\end{aligned}
\end{equation}
To achieve the maximal quantum violation, $4\sqrt{3}\approx 6.9282$, a choice for Alice's observables and the shared state is $\{A_1^*,A_2^*,A_3^*\}=\{\hat{Z},\hat{X},\hat{Y}\}$ and $\ket{\psi^*}^\text{AB}=\frac{1}{\sqrt{2}}(\ket{00}+\ket{11})$. By this we can define the \emph{elegant-Bell-type assemblage} as $\sigma_{a|x}^*=\psaxstar\ket{\hat{\sigma}^*_{a|x}}\bra{\hat{\sigma}^*_{a|x}}$ with $\psaxstar=1/2$ for all $a,x$ and
\begin{equation}
\begin{aligned}
&|\hat{\sigma}_{1|1}^*\rangle=\ket{0},~|\hat{\sigma}_{1|2}^*\rangle=\frac{\ket{0}+\ket{1}}{\sqrt{2}},~ {\blu |\hat{\sigma}_{1|3}^*\rangle=\frac{\ket{0}-i\ket{1}}{\sqrt{2}} },\\
&|\hat{\sigma}_{2|1}^*\rangle=\ket{1},~|\hat{\sigma}_{2|2}^*\rangle=\frac{\ket{0}-\ket{1}}{\sqrt{2}},~ {\blu |\hat{\sigma}_{2|3}^*\rangle=\frac{\ket{0}+i\ket{1}}{\sqrt{2}} },
\end{aligned}
\label{Eq_EB_assemblage}
\end{equation}
i.e., $\ket{\hatsaxstar}$ are the eigenstates of the three Pauli matrices. The representation of the elegant-Bell-type assemblage is shown in Fig.~\ref{Fig_EB_and_I3622}(a). In order to achieve the quantum bound of $I_{\rm E}$, Bob's observables can be chosen as $\{B_1^*,B_2^*,B_3^*,B_4^*\}=\{\frac{1}{\sqrt{3}}(\hat{Z}+\hat{X}-\hat{Y}),\frac{1}{\sqrt{3}}(\hat{Z}-\hat{X}+\hat{Y}),\frac{1}{\sqrt{3}}(-\hat{Z}+\hat{X}+\hat{Y}),\frac{1}{\sqrt{3}}(-\hat{Z}-\hat{X}-\hat{Y})\}$. Geometrically, the eigenstates of these observables form a regular tetrahedron on the Bloch sphere.

%

{\blubb
From Definition \ref{Def_2}, we need to use the controlled mixture $\sigma_{a|x}^{**}:=q\sigma_{a|x}^*\otimes|0\rangle\langle 0| + (1-q) (\sigma_{a|x}^*)^\mathsf{T}\otimes|1\rangle\langle 1|$ as the reference assemblage. Bob's optimal observables are accordingly $B_y^{**}=B_y^*\otimes |0\rangle\langle 0|+(B_y^*)^\mathsf{T}\otimes |1\rangle\langle 1|$. Following the same procedure of the previous cases, i.e., the grey box below Eq.~\eqref{Eq_classical_fidelity_SDP}, and with a little bit involved computation, we can derive a DI expression of the fidelity $\mathcal{F}$. See Appendix \ref{SecApp_EBIandI3622} for more detail. A lower bound on $\mathcal{F}$ for a quantum violation of $I_{\rm E}$ can then be computed by the following SDP (similarly for the case of $I_{3622}$):
\begin{equation}
\begin{aligned}
\min\quad &\mathcal{F}[\bm{\sigma}^{**},\Lambda(\bm{\rho})]\\
\text{s.t.} \quad &I_{\rm E}(\mathbf{P}) = I_{\rm E}^\text{obs}\\
& \chi[\sax,\mathcal{S}]\succeq 0,\\
& P_{\bm{\rho}}(a|x)=P_{\bm{\rho}}^{\rm obs}(a|x),\\
& \sum_a\chi[\sax,\mathcal{S}] = \sum_a\chi[\sigma_{a|x'},\mathcal{S}],\\
&\chi_\text{\tiny L}[\sax,B_5(B_1+B_2-B_3-B_4),\mathcal{S}']\succeq 0,\\
&\chi_\text{\tiny L}[\sax,B_6(B_1-B_2+B_3-B_4),\mathcal{S}']\succeq 0,\\
&\chi_\text{\tiny L}[\sax,B_7(-B_1+B_2+B_3-B_4),\mathcal{S}']\succeq 0.
\end{aligned}
\label{Eq_min_F_SDP_EBI}
\end{equation}
}

{\blubb The derivation of the constraints related to localizing matrices, i.e., the last three constraints, is shown in Appendix~\ref{SecApp_EBIandI3622}. The results are plotted in Fig.~\ref{Fig_EB_and_I3622}(b), therein the $x$-axis represents for the percentage of deviation from the maximal quantum violation of $I_{\rm E}$.  The threshold is around $5.4\%$ away from the maximal quantum violation.
}

\section{Applications}\label{Sec_applications}
\subsection{DI certification of all entangled two-qubit states }\label{Sec_DI_all_ent}
\begin{figure}[h]
\includegraphics[width=1\linewidth]{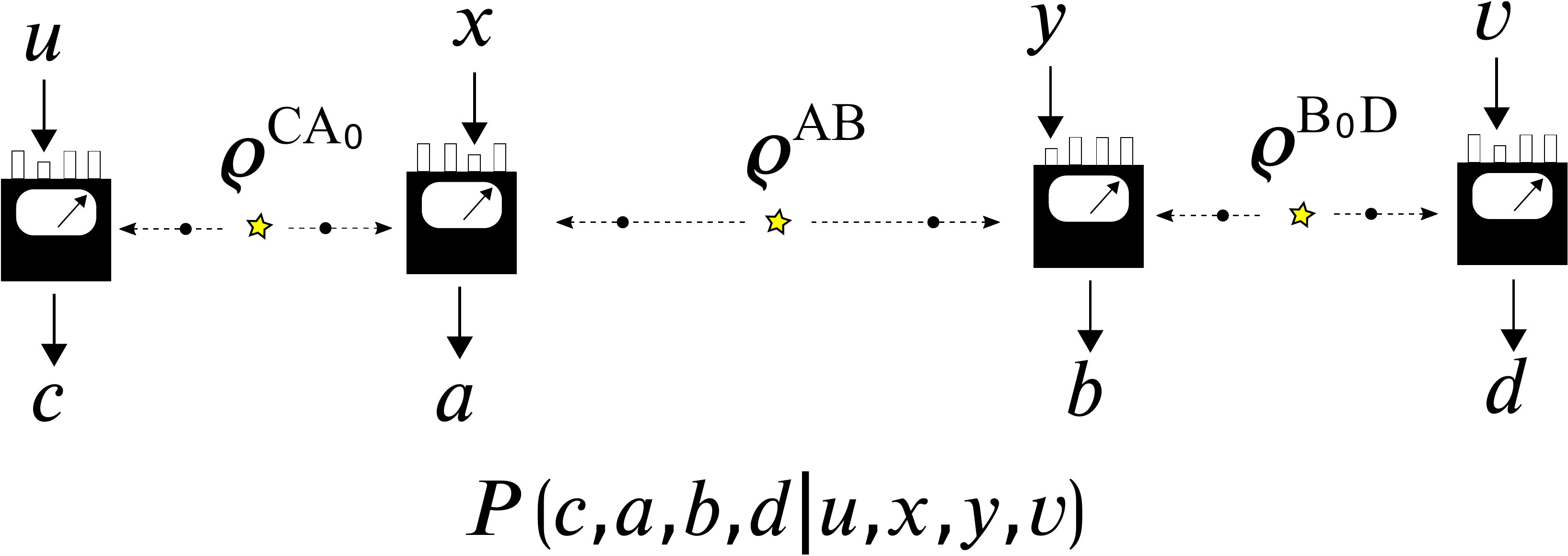}
\caption{The protocol for DI certification of all entangled states proposed by Bowles \emph{et al.}~\cite{Bowles18,Bowles18PRA}. To verify entanglement of the quantum state $\rab$, two more parties, Charlie and Daisy, join the certification task and distribute quantum states $\rca$ and $\rbd$ with, respectively, Alice's and Bob's auxiliaries. In this network composed of four black boxes, three correlations will be observed, i.e., $\{P(c,a|u,x)\}$, $\{P(b,d|y,v)\}$, and $\{P(c,a,b,d|u,x,y,v)\}$. If both $\{P(c,a|u,x)\}$ and $\{P(b,d|y,v)\}$ successful self-test Chalire's measurements, $\rca$, $\rbd$, and Daisy's measurements, then one can find a DI entanglement witness for $\{P(c,a,b,d|u,x,y,v)\}$ caused by any entangled state $\rab$.
}
\label{Fig_CABD}
\end{figure}

It is well known that there exist entangled states admitting a local-hidden-variable model~\cite{Werner89}. In other words, if Alice and Bob would like to verify entanglement of their share $\rab$ through the observed correlation $\{P(a,b|x,y)\}$ violating a Bell inequality, some entangled states will fail to be detected. In 2012, Buscemi~\cite{Buscemi12} showed that by further introducing two tomographically complete sets of states $\{\hat{\tau}_x\}$ and $\{\hat{\omega}_y\}$, respectively, for Alice and Bob, all entangled states can be certified through $\{P(a,b|x,y)\}$, though not in a fully DI manner (see \cite{Cavalcanti13} for verifying all steerable states and \cite{Zhao2020} for verifying all steerable assemblages). Recently, Bowles \emph{et al.}~\cite{Bowles18,Bowles18PRA} considered to introduce two more parties, called Charlie and Daisy (see Fig.~\ref{Fig_CABD}), who share quantum states $\rca\in\mathsf{L}(\mathcal{H}_\text{C}\otimes\mathcal{H}_{\text{A}_0})$ and $\rbd\in\mathsf{L}(\mathcal{H}_{\text{B}_0}\otimes\mathcal{H}_{\text{D}})$, respectively, with Alice's and Bob's auxiliaries. Charlie (Daisy) also performs measurement on his (her) share in a black box scenario, i.e., he (she) obtains a collection of outcomes $\{c\}$ ($\{d\}$) when performing measurements labelled from the set $\{u\}$ ($\{v\}$). At the end of experiment, the following statistics are obtained: $\{P(c,a|u,x)\}$, $\{P(b,d|y,v)\}$, and $\{P(c,a,b,d|u,x,y,v)\}$. The first (second) correlation is used for self-testing that Charlie's (Daisy's) measurements and the share state $\rca$ ($\rbd$) are indeed in the reference ones, ensuring that the set of states steered on Alice's (Bob's) part, i.e., $\{\hattcu\}$ ($\{\hatodv\}$), is the ideal one. The third correlation is used for verifying the entangled state $\rab$ through the DI entanglement witness $\vec{\beta}\cdot\mathbf{P}:=\sum_{c,a,b,d,u,x,y,v}\beta_{c,a,b,d}^{u,x,y,v}P(c,a,b,d|u,x,y,v)\geq 0$, with $\vec{\beta}$ being a set of  real numbers.

With our robust self-testing method, we would like to show that it also allows one to provide the same qualitative result for the shared state being two-qubit entangled state.
The idea behind is that we self-test the reference steered assemblages $\{\tcu\}$ and $\{\odv\}$ (hence $\{\hattcu\}$ and $\{\hatodv\}$ ) instead of self-testing the measurements (of Charlie and Daisy) and the states $\rca$, $\rbd$. In what follows we will show two facts. The first one is that when both the self-testing of Alice's and Bob's assemblages (prepared by Charlie and Daisy, respectively) are perfectly achieved, for any given entangled qubit state $\rab$ there exists a DI witness $\vec{\beta}$ such that $\vec{\beta}\cdot \mathbf{P} < 0$, while $\vec{\beta}\cdot \mathbf{P}\geq 0$ for all separable states. The other one is that when the self-testing is imperfect (i.e., with the fidelity deviating from the value of $1$), one is able to shift the separable bound $0$ to avoid detecting separable states, though some entangled states cannot be verified. {\blu Note that to arrive the property of positivity of the witness, {\blub namely $\vec{\beta}\cdot \mathbf{P}\geq 0$}, we do not make any assumption on $\rab$, including its dimension.}

First, consider that Charlie and Alice use the $I_{3622}$ inequality to perform the self-testing:\footnote{Since the part of Bob and Daisy is the same as Charlie and Alice, we only discuss the latter. The self-testing part of the former can be trivially obtained.}~\cite{Acin16,Bowles18,Bowles18PRA}
\begin{equation}
\begin{aligned}
I_{3622}^{\rm {\tiny CA}}&:=\langle C_1A_1\rangle + \langle C_1A_2\rangle + \langle C_2A_1\rangle - \langle C_2A_2\rangle\\
&+\langle C_1A_3\rangle + \langle C_1A_4\rangle - \langle C_3A_3\rangle + \langle C_3A_4\rangle\\
&+\langle C_2A_5\rangle + \langle C_2A_6\rangle - \langle C_3A_5\rangle + \langle C_3A_6\rangle\stackrel{\mathcal{L}}{\leq} 6,
\end{aligned}
\label{Eq_I3622_CA}
\end{equation}
where $\langle C_uA_x\rangle:=P(c=a|u,x)-P(c\neq a|u,x)$. As it has been shown in Section \ref{Sec_EB_type}, the maximal quantum violation of this inequality self-tests that Alice's assemblage $\{\tcustar\}$ is of the form of Eq.~\eqref{Eq_EB_assemblage}, which is tomographically complete.

Having successfully self-tested Alice's assemblage (prepared by Charlie) and Bob's assemblage (prepared by Daisy), the final step is to construct a DI entanglement witness $I_{\rm DIEW}:=\vec{\beta}\cdot\mathbf{P}$ and show how it can be used for certifying the entangled state $\rab$. The DI entanglement witness we use is the same as that of Ref.~\cite{Bowles18,Bowles18PRA}:
\begin{equation}
I_{\rm DIEW}:=\sum_{c,d,u,v}\beta_{c,d}^{u,v} P(c,+,+,d|u,\blacklozenge,\blacklozenge,v),
\label{Eq_DIEW}
\end{equation}
where $x=\blacklozenge$ and $y=\blacklozenge$, respectively, represent for Alice's and Bob's $7$th measurement settings. In Appendix \ref{SecApp_DIEW}, we show $I_{\rm DIEW}$ is capable of certifying all entangled two-qubit states when both self-testing of Alice's and Bob's assemblages are perfect (i.e., $I_{3622}=6\sqrt{2}$). If the self-testing is imperfect, i.e., $I_{3622}$ departs from the value of $6\sqrt{2}$, we also show the separable bound of $I_{\rm DIEW}$ can be shifted to avoid wrongly detecting separable states, though some entangled states will fail to be detected in this case. {\blu To verify higher dimensional entangled state, we need to construct a framework of \emph{parallel self-testing} with our method. This issue is beyond the scope of our present work and we leave it for future research.} Note that most of the techniques we use are the same as those of Refs.~\cite{Bowles18,Bowles18PRA}, therefore we leave the detail of the proof in Appendix \ref{SecApp_DIEW}.

{\blu As pointed out in Introduction, given that the standard self-testing of the measurements and the state have been derived for some Bell scenarios, one may suggest to use the idea of error propagation to derive bounds on the fidelity. Nevertheless, an advantage of our method is that it gives better robustness than the standard self-testing (at least, for the scenarios considered in this work). For the $I_{3622}$ scenario, the tolerance that Ref.~\cite{Bowles18PRA} gives is around $10^{-3}\%$, i.e., no self-testing statement can be drawn if the violation of $I_{3622}$ departs $10^{-3}\%$ away from the maximal quantum violation. By contrast, the tolerance that our method gives is around {\blubb $5.4\%$ (see Fig.~\ref{Fig_DI_fidelity_I3622_contr_mix} in Appendix \ref{SecApp_EBIandI3622}).} However, we would like to point out that it may not make much sense to compare the robustness in this entanglement certification task. The reason is that if the violation of $I_{3622}$ departs from the maximal quantum violation, one has to check, in a DI way, if Alice's assemblage $\{\tcu\}$ (prepared by Charlie) and Bob's assemblage $\{\odv\}$ (prepared by Daisy) remain tomographically complete, which is still an open question.
}

\subsection{DI certification of all non-entanglement-breaking qubit channels }\label{Sec_DI_all_qchannel}
\begin{figure}
\begin{minipage}[c]{.49\textwidth}
\includegraphics[width=7.5cm]{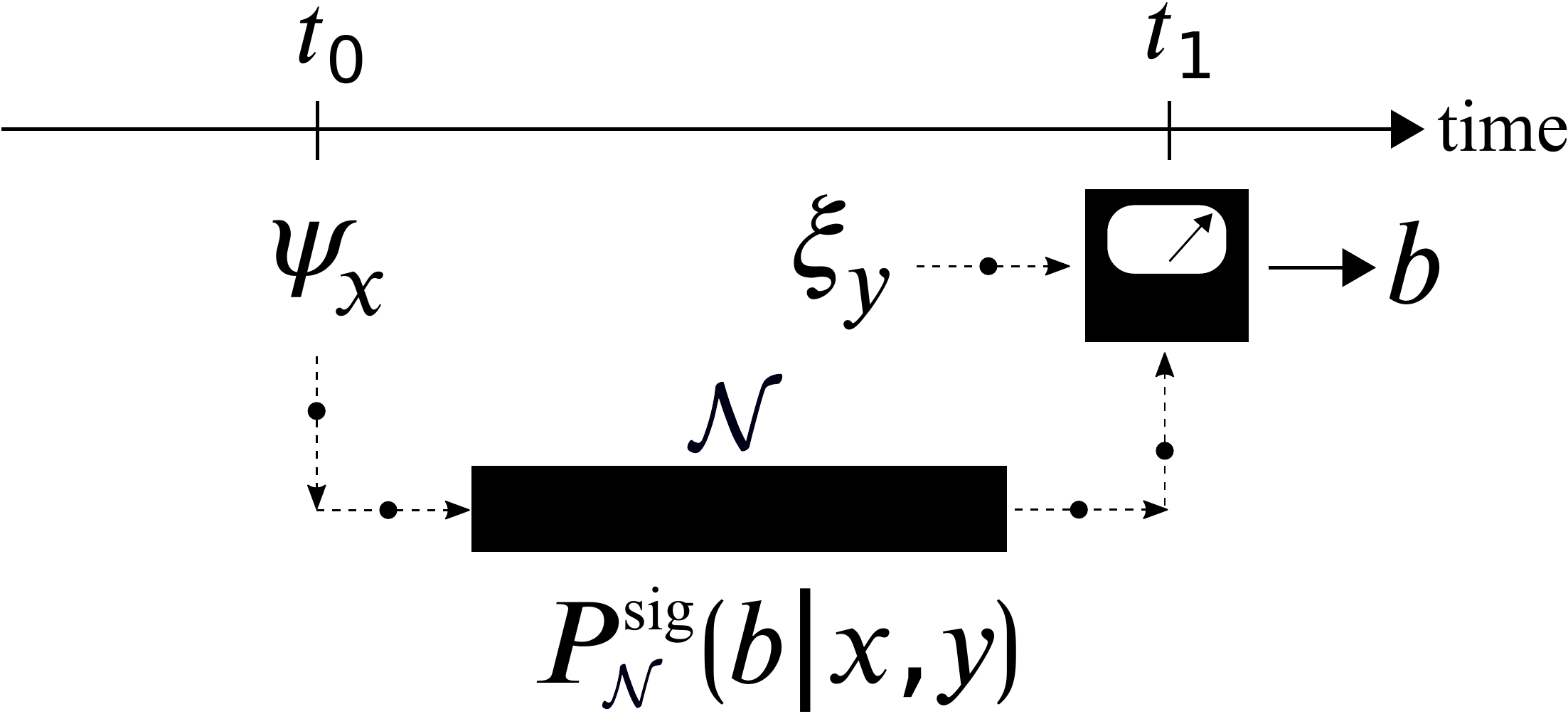} \\
\centering
\text{(a)}
\end{minipage}
\begin{minipage}[c]{.49\textwidth}
\includegraphics[width=7.5cm]{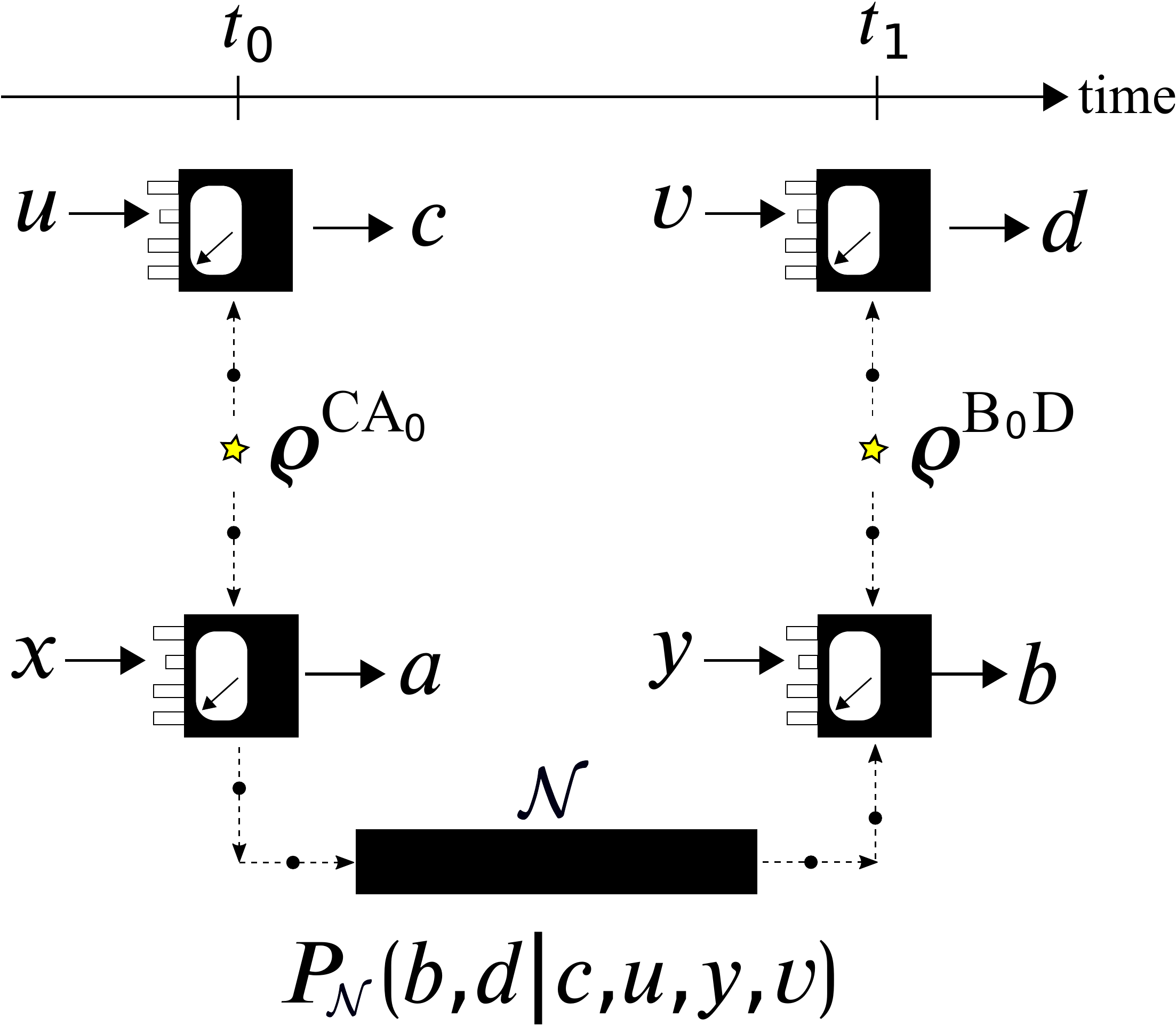}\\
\centering
\text{(b)}
\end{minipage}
\caption{ (a) The protocol for semi-quantum signalling games~\cite{Rosset18}. At time $t_0$, a characterized quantum state $\psi_x\in\{\psi_x\}$ is sent into a quantum channel $\mathcal{N}$. After the evolution, the involved state $\mathcal{N}(\psi_x)$ and another characterized state $\xi_y\in\{\xi_y\}$ are jointly measured by a black box. Under such a framework, Ref.~\cite{Rosset18} showed that any non-entanglement-breaking (non-EB) channel can be verified by properly choosing a witness. (b) The protocol for DI certification of all non-EB channels. To verify that channel $\mathcal{N}$ is non-EB in a DI scheme, two black boxes are used at time $t_0$ to obtain the correlation $\{P(c,a|u,x)\}$ for self-testing of Alice's reference assemblage (c.f. $\psi_x$ in (a)). The usage of the other two black boxes at time $t_1$ is to obtain the correlation $\{P(b,d|y,v)\}$ for self-testing of Bob's reference assemblage (c.f. $\xi_y$ in (a)). If both of the self-testing are successful, one can find witnesses for correlations $\{P_{\mathcal{N}}^{\rm DI}(b,d|c,u,y,v)\}$ induced by any non-EB quantum channel.
}
\label{Fig_signaling_game_DI}
\end{figure}

As the second application, we would like to show that our tool can also be used for certifying all non-entanglement-breaking (non-EB) qubit channels. In Ref.~\cite{Rosset18}, Rosset \emph{et al.} proposed the so-called \emph{semi-quantum signalling games} and showed that under such a framework, any non-EB channel outperforms EB channels. More specifically, in a semi-quantum signalling game, a well-characterized quantum state $\ket{\psi_x}\in\{\ket{\psi_x}\}_x$ is prepared by Alice and sent into a quantum channel $\mathcal{N}$ at time $t_0$. After some time, the state evolves into $\mathcal{N}(\ket{\psi_x})$ and is measured by Bob jointly with another state $\xi_y\in\{\ket{\xi_y}\}_y$ at time $t_1>t_0$. See Fig.~\ref{Fig_signaling_game_DI}(a) for the schematic diagram. After many rounds, one obtains a set of probability distributions $\{P_{\mathcal{N}}^{\rm sig}(b|x,y)\}$. If $\{\ket{\psi_x}\}$ and $\{\ket{\xi_y}\}$ respectively form tomographically complete sets, one can construct a witness $I_\mathcal{N}^{\rm sig}:=\sum_{b,x,y}\gamma_{b,x,y}P_{\mathcal{N}}^{\rm sig}(b|x,y)$ such that  $I_\mathcal{N}^{\rm sig} < 0$ for the given non-EB channel while $I_\mathcal{N}^{\rm sig} \geq 0$ for all EB channels~\cite{Rosset18}. In what follows, we show how to remove the necessity of the characterization of $\{\psi_x\}$ and $\{\xi_y\}$ therefore the entire framework will be device-independent.

The idea is similar to the one used in the works of Bowles \emph{et al.}~\cite{Bowles18,Bowles18PRA}. For simplicity, we consider that the non-EB channel we are going to verify is a qubit channel. Then, we introduce two more parties, called Charlie and Daisy, who share the maximally entangled states with Alice and Bob, respectively, and perform the Pauli measurements $\hat{Z}$, $\hat{X}$, $\hat{Y}$ on their shares. By doing so, each of Alice and Bob obtains a set of eigenstates of the Pauli observables, which is tomographically complete. As shown in the previous sections, we are indeed able to self-test the assemblage associated with this set of states by the maximal quantum violation of either $I_{3622}$ or $I_{\rm E}$. The entire DI setting is depicted in Fig.~\ref{Fig_signaling_game_DI}(b). As can be seen, at time $t_0$, one performs self-testing of Alice's (the black box with action $(a,x)$) reference assemblage $\{\tcustar\}$ prepared by Charlie (the black box with action $(c,u)$) by the observation of the maximal quantum violation of, say $I_{\rm E}$. Then, the state $\hattcustar$ is sent into the quantum channel. At time $t_1$, one performs the other self-testing of Bob's reference assemblage $\{\odvstar\}$ prepared by Daisy. Besides, the evolved state $\mathcal{N}(\hattcustar)$ is jointly measured with $\hatodvstar$ and the set of probability distributions $\mathbf{P}_{\mathcal{N}}:=\{P_{\mathcal{N}}(b=+,d|c,u,y=\blacklozenge,v)\}$ is obtained, where $\blacklozenge$ is the $5$th measurement setting of $y$. In Appendix~\ref{SecApp_DI_nonEBC}, we show that for a correlation $\mathbf{P}_{\mathcal{N}}$ caused by any non-EB qubit channel, there exists a set of coefficients $\{\gamma_{c,d}^{u,v}\}$, namely a witness, such that
\begin{equation}
\sum_{c,d,u,v}\gamma_{c,d}^{u,v} P_{\mathcal{N}}(+,d|c,u,\blacklozenge,v) < 0
\label{Eq_DI_nonEBC}
\end{equation}
while $\sum_{c,d,u,v}\gamma_{c,d}^{u,v} P_{\mathcal{N}}^{\rm EB}(+,d|c,u,\blacklozenge,v)\geq 0$ for correlations $\mathbf{P}_{\mathcal{N}}^{\rm EB}$ caused by all EB channels.


\section{Discussion}\label{Sec_discussion}

In this work, we introduce the framework of \emph{robust self-testing of steerable quantum assemblages}, which provides quantitative estimation of how close the underlying assemblage is to the reference one when the Bell inequality may not achieve the maximal quantum violation. The framework is device-independent (DI), i.e., no assumption is made on the measurements involved nor on the underlying state shared between Alice and Bob. We give several types of self-testable assemblages such as the CHSH type, the tilted CHSH type, the elegant Bell type, and the $I_{3622}$ type. Fundamentally, this work classifies different types of steerable quantum assemblages and explores the relation between these assemblages and the boundary of the quantum set of correlations. We also give two explicit applications on DI quantum certification: 1) It can be used for an alternative proof of the protocol of Ref.~\cite{Bowles18}, i.e., DI certification of all entangled states, and 2) it can be used for constructing a DI certification of all non-entanglement-breaking channels, which is with fewer assumptions compared with the work of Ref.~\cite{Rosset18}.

We would like to point out that although an assemblage is produced by performing a set of measurements on a shared state, the successful self-testing of an assemblage does \emph{not necessarily} imply the successful self-testing of the state or measurements. This is due to the fact that a single assemblage itself cannot fully characterize the measurements and state that are used to generate it. Our work therefore poses the following open question: Is it always true that if a given assemblage is self-testable, one can, in addition, self-test the associated measurements and state?



We also leave some open problems in the following. First, we do not follow specific rules to express the CJ matrix corresponding to the identity channel (i.e., the unnormalized maximally entangled state) in terms of Bob's reference measurements. Is there a general way to express the maximally entangled state in terms of the reference measurements, so that it could make the entire framework more universal? Second, compared with the DI bounds on fidelities obtained in this work, are there better strategies, i.e., better expressions of the maximally entangled state, that give greater bounds? Third, in our application sections, we only consider that each party holds a qubit system in the DI verification task. It is expected that the numerical computations of higher dimensional states will be considered in future research. Fourth, we look forward to an analytical frameworks of the robust self-testing of assemblages, using either the typical method based on the trace distance~\cite{McKague12} or the method based on operator inequalities~\cite{Kaniewski16,Coopmans19}. Finally, our work may be generalized to multipartite scenarios, i.e., certification of assemblages in a multipartite setting by observing the maximal quantum violation of a multipartite Bell inequality.

\begin{acknowledgments}
We acknowledge Daniel Cavalcanti, Jebarathinam Chellasamy, J{\k{e}}drzej Kaniewski, Yeong-Cherng Liang, and Nathan Walk for the useful discussion. {\blu We also thank the anonymous referees for very helpful comments.} This work was funded by the Deutsche Forschungsgemeinschaft (DFG, German Research Foundation) - Project number 414325145 in the framework of the Austrian Science Fund (FWF): SFB F71.
SLC acknowledges the support of the Ministry of Science and Technology, Taiwan (MOST Grants No. 107-2917-I-564-007, No. 108-2811-M-006-515, No. 109-2811-M-006 -509, and No. 110-2811-M-006-539). HYK acknowledges the support of the Ministry of Science and Technology, Taiwan (MOST Grants No. 108-2811-M-006-536, No. 109-2811-M-006-516, and No. 110-2811-M-006- 546). WZ is supported by the Program (RWDC) for Leading Graduate Schools of Nagoya University. JT thanks the Alexander von Humboldt foundation for support. YNC acknowledges the support of the Ministry of Science and Technology, Taiwan (MOST Grants No. 107-2628-M-006-002-MY3, No.  109-2627-M-006-004, and No. 110-2123-M-006-001), and the U.S. Army Research Office (ARO Grant No. W911NF-19-1-0081).
\end{acknowledgments}

\clearpage
\onecolumngrid
\appendix
\section{Classical fidelity obtained from LHS model}\label{SecApp_classical_fidelity}
In this section we derive the SDP computing the trivial fidelity of given reference assemblage, i.e., Eq.~\eqref{Eq_classical_fidelity_SDP}
\begin{equation}
\begin{aligned}
\max_{\{\hat{\varsigma}_\lambda\}} \quad & \frac{\sqrt{|\mathcal{A}|}}{|\mathcal{X}||\lambda|}\sum_{a,x,\lambda}\sqrt{\psaxstar} \delta_{a,\lambda_x}\tr(\hatsaxstar\hat{\varsigma})\\
\text{such that}\quad & \hat{\varsigma}_\lambda\succeq 0,\quad \tr(\hat{\varsigma}_\lambda)=1 \quad\forall\lambda.
\end{aligned}
\label{EqApp_classical_fidelity_SDP}
\end{equation}
Recall that in Sec.~\ref{Sec_steering} a local-hidden-state model of assemblages is written as~\cite{Wiseman07}
\begin{equation}
\rho_{a|x}^{\rm US} = \sum_\lambda P(\lambda)\delta_{a,\lambda_x}\hat{\varsigma}_\lambda\quad\forall a,x,
\end{equation}
where $\hat{\varsigma}_\lambda$ are normalized quantum states for all $\lambda$. The superscript ``US'' denotes for ``unsteerable''. Without loss of generality, we assume that the probability $P(\lambda)$ used for distributing the classical strategy is uniform, i.e., $P(\lambda)=1/|\lambda|$ with $|\lambda|=|\mathcal{A}|^{|\mathcal{X}|}$ being the number of possible vectors $\lambda$. Therefore for all $a$ and $x$ we have
\begin{equation}
P_{\bm{\rho}}(a|x):=\tr(\rho_{a|x}^{\rm US}) = \frac{1}{|\lambda|}\sum_\lambda\delta_{a,\lambda_x} = \frac{1}{|\lambda|}\frac{|\lambda|}{|\mathcal{A}|} = \frac{1}{|\mathcal{A}|},
\end{equation}
where in the third equality we use the fact that the number of the non-zero elements of the set $\{\delta_{a,\lambda_x}\}_{\lambda}$ is $|\mathcal{A}|^{|\mathcal{X}|-1}$ for all $a$ and $x$. With this, given a reference assemblage $\bm{\sigma}^*$, the best fidelity that $\bm{\rho}^{\rm US}$ can achieve is
\begin{equation}
\begin{aligned}
f_{\rm c}
&:= \max_{\bm{\rho}^{\rm US}} \mathcal{F}(\bm{\sigma},\bm{\rho})\\
&:= \max_{\bm{\rho}^{\rm US}} \sum_{a,x}\sqrt{\frac{P^*_{\bm{\sigma}}(a|x)}{P_{\bm{\rho}}(a|x)}} \bra{\hatsaxstar}\rho_{a|x}^{\rm US}\ket{\hatsaxstar}\\
&=\max_{\hat{\varsigma}_{\lambda}}\frac{1}{|\mathcal{X}|}\sum_{a,x,\lambda}\sqrt{|\mathcal{A}|}\sqrt{P^*_{\bm{\sigma}}(a|x)}\bra{\hatsaxstar} P(\lambda)\delta_{a,\lambda_x}\hat{\varsigma}\ket{\hatsaxstar}\\
&=\max_{\hat{\varsigma}_{\lambda}}\frac{\sqrt{|\mathcal{A}|}}{|\mathcal{X}||\lambda|}\sum_{a,x,\lambda}\sqrt{P^*_{\bm{\sigma}}(a|x)}\delta_{a,\lambda_x}\tr(\hatsaxstar\hat{\varsigma}).
\end{aligned}
\end{equation}

\section{SDP computing DI lower bounds on the fidelity of the elegant-Bell-type and $I_{3622}$-type assemblages}\label{SecApp_EBIandI3622}
\subsection{The elegant Bell scenario}
{\blubb In this section, we provide the detail derivation of the SDP~\eqref{Eq_min_F_SDP_EBI}, including the fidelity expression $\mathcal{F}$ of the objective function. To this end, recall that to self-test complex-valued assemblages, we have to compute lower bounds on the fidelity between the underlying assemblage and the controlled mixture of the reference assemblage and its transpose (see Def.~\ref{Def_2}), i.e., a lower bound on $\mathcal{F}[\bm{\sigma}^{**},\Lambda(\bm{\rho})]$, where $\sigma_{a|x}^{**}:=q\sigma_{a|x}^*\otimes|0\rangle\langle 0| + (1-q) (\sigma_{a|x}^*)^\mathsf{T}\otimes|1\rangle\langle 1|$. Unlike the situation of self-testing real-valued assemblage, now $\hat{\sigma}_{a|x}^{**}:=\sigma_{a|x}^{**}/\tr(\sigma_{a|x}^{**})$ are not pure states any more. Therefore we use the so-called $p$-fidelity (which is based on the Schatten $p$-norm) with $p=2$~\cite{Liang19}, denoted as $F_2$, as a measure between two quantum states. As such, the fidelity between two assemblages can be generalized as (cf. Eq.~\eqref{Eq_assemblage_fidelity})
\begin{equation}
\begin{aligned}
&\mathcal{F}(\bm{\sigma},\bm{\rho})=\frac{1}{|\mathcal{X}|}\sum_{a,x} \sqrt{\psax\prax}~F_2(\hatsax,\hatrax),\\
\end{aligned}
\end{equation}
where
\begin{equation}
F_2(\hatsax,\hatrax):=\frac{\tr(\hatsax\hatrax)}{\max [\tr(\hat{\sigma}_{a|x}^2),\tr(\hat{\rho}_{a|x}^2)]}.
\end{equation}
Through this, the term $\mathcal{F}[\bm{\sigma}^{**},\Lambda(\bm{\rho})]$ can then be written as
\begin{equation}
\mathcal{F}[\bm{\sigma}^{**},\Lambda(\bm{\rho})]=\frac{1}{|\mathcal{X}|}\sum_{a,x} {\sqrt{\psaxstarstar\prax}}\frac{\tr[\hat{\sigma}_{a|x}^{**}\Lambda(\hat{\rho}_{a|x})]}{\max \Big[\tr[(\hat{\sigma}_{a|x}^{**})^2],\tr[\Lambda(\hat{\rho}_{a|x})^2]\Big]}.
\label{EqApp_fidelity2}
\end{equation}
In the above equation,  {\bluiv $\Lambda(\hat{\rho}_{a|x})$ can be expanded as $\Lambda(\hat{\rho}_{a|x})=\sum_{ijkl}C_{aij}^{xkl}\ket{i}\bra{j}\otimes\ket{k}\bra{l}$, where $C_{aij}^{xkl}$ are some numbers. We then apply a dephasing map, which is CP, on $\Lambda(\hat{\rho}_{a|x})$ and eliminate the $\ket{0}\bra{1}$, $\ket{1}\bra{0}$ terms of the auxiliaries of $\Lambda(\hat{\rho}_{a|x})$. As such, we can write $\Lambda(\hat{\rho}_{a|x})$ in the form of {\bluv $p_{a|x}\hat{\rho}'_{a|x}\otimes |0\rangle\langle 0|+ (1-p_{a|x})\hat{\rho}''_{a|x}\otimes |1\rangle\langle 1|$, with $0\leq p_{a|x}\leq 1$ and $\hat{\rho}'_{a|x},\hat{\rho}''_{a|x}$ being some states associated with some assemblages. Choosing $q$ of $\hat{\sigma}_{a|x}^{**}$ to be equal to $\max\{ \{p_{a|x}, 1-p_{a|x}\}_{a,x}\}$,  we have $\tr[\Lambda({\hat{\rho}}_{a|x})^2]=p_{a|x}^2\tr[(\hat{\rho}'_{a|x})^2]+(1-p_{a|x})^2\tr[(\hat{\rho}''_{a|x})^2]\leq p_{a|x}^2 + (1-p_{a|x})^2 \leq q^2+(1-q)^2=\tr[(\hat{\sigma}_{a|x}^{**})^2]$. }Consequently, we have
\begin{equation}
\mathcal{F}[\bm{\sigma}^{**},\Lambda(\bm{\rho})]=\frac{1}{|\mathcal{X}|}\sum_{a,x}\frac{1}{q^2+(1-q)^2} \frac{1}{\sqrt{\psaxstarstar\prax}}~\tr[\sigma_{a|x}^{**}\Lambda(\rho_{a|x})].
\label{EqApp_fidelity3}
\end{equation}


}


To achieve the maximal quantum violation of the elegant Bell inequality, Bob can measure the observables: $B_y^{**}=B_y^*\otimes |0\rangle\langle 0| + (B_y^*)^\mathsf{T}\otimes |1\rangle\langle 1|$, where
\begin{equation}
\begin{aligned}
&B_1^*=\frac{1}{\sqrt{3}}(\hat{Z}+\hat{X}-\hat{Y}),\quad B_2^*=\frac{1}{\sqrt{3}}(\hat{Z}-\hat{X}+\hat{Y}),\\
&B_3^*=\frac{1}{\sqrt{3}}(-\hat{Z}+\hat{X}+\hat{Y}),\quad B_4^*=\frac{1}{\sqrt{3}}(-\hat{Z}-\hat{X}-\hat{Y}).
\end{aligned}
\label{EqApp_optBy}
\end{equation}
As in the tilted-CHSH-type scenario in the main text, here, we also have to introduce observables $B_5^{**}$, $B_6^{**}$, and $B_7^{**}$, such that
\begin{equation}
\begin{aligned}
&B_5^{**}\left[\frac{\sqrt{3}}{4}\Big(B_1^{**}+B_2^{**}-B_3^{**}-B_4^{**}\Big)\right]\succeq 0,\\
&B_6^{**}\left[\frac{\sqrt{3}}{4}\Big(B_1^{**}-B_2^{**}+B_3^{**}-B_4^{**}\Big)\right]\succeq 0,\\
&B_7^{**}\left[\frac{\sqrt{3}}{4}\Big(-B_1^{**}+B_2^{**}+B_3^{**}-B_4^{**}\Big)\right]\succeq 0.
\end{aligned}
\label{EqApp_PolarDecomp}
\end{equation}
Since the entire term in each pair of square brackets is unitary, we have $B_5^{**}=\hat{Z}\otimes\openone$, $B_6^{**}=\hat{X}\otimes\openone$, and $B_7^{**}=\hat{Y}\otimes\hat{Z}$.

The procedure of representing the CJ matrix is a little bit involved. First, we clearly write down the Hilbert space that the reference assemblage and the underlying assemblage lives in: $\sigma_{a|x}^{**}\in\mathsf{L}(\mathcal{H}_{{\rm B}'}\otimes\mathcal{H}_{{\rm B}''})$ and $\rho_{a|x}\in\mathsf{L}(\mathcal{H}_{\rm B})=\mathsf{L}(\mathcal{H}_{{\rm B}_1}\otimes\mathcal{H}_{{\rm B}_2})$. Therefore we have $\Lambda: \mathsf{L}(\mathcal{H}_{{\rm B}_1}\otimes\mathcal{H}_{{\rm B}_2})\rightarrow\mathsf{L}(\mathcal{H}_{{\rm B}'}\otimes\mathcal{H}_{{\rm B}''})$. When performing the characterized optimal quantum strategy, we have $\bm{\rho}=\bm{\sigma}^*$, therefore we again choose $\Lambda$ as the identity map, i.e., $\Lambda={\rm id}\otimes {\rm id}=\Lambda^{{{\rm B}_1}\rightarrow {\rm B}'}\otimes\Lambda^{{{\rm B}_2}\rightarrow {\rm B}''}$. The corresponding CJ matrix $\Omega$ will be the Kronecker product of two (unnormalized) maximally entangled states: 
\begin{equation}
\Omega=\Omega_1^{{{\rm B}_1}{\rm B}'}\otimes\Omega_2^{{{\rm B}_2}{\rm B}''}=\left(\sum_{ij}|i\rangle^{{\rm B}_1}\langle j|\otimes|i\rangle^{{\rm B}'}\langle j|\right)\otimes\left(\sum_{kl}|k\rangle^{{\rm B}_2}\langle l|\otimes|k\rangle^{{\rm B}''}\langle l|\right).
\end{equation}
We then obtain
\begin{equation}
[\Lambda(\rho_{a|x}^{\rm B})]^\mathsf{T}=\tr_{\rm B} \Big[ \Big(\rho_{a|x}^{\rm B} \otimes\openone^{{\rm B}'{\rm B}''}\Big) \Omega^{\mathsf{T}}\Big]=\sum_{ijkl}\tr\Big[ \rho_{a|x}^{\rm B}\Big(|i\rangle^{{\rm B}_1}\langle j|\otimes |k\rangle^{{\rm B}_2}\langle l|\Big) \Big] |i\rangle^{{\rm B}'}\langle j|\otimes |k\rangle^{{\rm B}''}\langle l|
\end{equation}
and hence\footnote{We omit the superscripts used for representing the Hilbert space when there is no risk of confusion.}
\begin{equation}
\begin{aligned}
\tr\Big[(\sigma_{a|x}^{**})^\mathsf{T}\big[ \Lambda(\rho_{a|x}) \big]^\mathsf{T}\Big]&=
q\cdot\sum_{ij}\tr\Big[ \rho_{a|x}\Big( |i\rangle\langle j|\otimes |0\rangle\langle 0| \Big) \Big]\cdot\langle j|(\sigma^*)^\mathsf{T}|i\rangle \\
&~~+
(1-q)\cdot\sum_{ij}\tr\Big[ \rho_{a|x}\Big( |i\rangle\langle j|\otimes |1\rangle\langle 1| \Big) \Big]\cdot\langle j|\sigma^*|i\rangle\\
&=q\cdot\tr\Big[\rho_{a|x}\Big(\sigma_{a|x}^*\otimes |0\rangle\langle 0|\Big)\Big] + (1-q)\cdot\tr\Big[\rho_{a|x}\Big((\sigma_{a|x}^*)^\mathsf{T}\otimes |1\rangle\langle 1|\Big)\Big].
\end{aligned}
\end{equation}
Inserting the above terms into Eq.~\eqref{EqApp_fidelity2}, we obtain an expression of the fidelity:\footnote{ Here we assume that the marginal $P_{\bm{\rho}}(a|x)=1/2$ is observed. For other values of $P_{\bm{\rho}}(a|x)$, one follows the same steps and obtains a variant form of the fidelity.}
\begin{equation}
\begin{aligned}
\mathcal{F}[\bm{\sigma}^{**},\Lambda(\bm{\rho})] &=  \frac{1}{3}\cdot\frac{1}{q^2+(1-q)^2}\cdot\\
\Big\{q
\Big\{
&\tr\Big[\rho_{1|1}\Big(\frac{\openone + \hat{Z}}{2}\otimes |0\rangle\langle 0|\Big)\Big]+
\tr\Big[\rho_{2|1}\Big(\frac{\openone - \hat{Z}}{2}\otimes |0\rangle\langle 0|\Big)\Big]
+\tr\Big[\rho_{1|2}\Big(\frac{\openone + \hat{X}}{2}\otimes |0\rangle\langle 0|\Big)\Big]\\
+&\tr\Big[\rho_{2|2}\Big(\frac{\openone - \hat{X}}{2}\otimes |0\rangle\langle 0|\Big)\Big]
+\tr\Big[\rho_{1|3}\Big(\frac{\openone - \hat{Y}}{2}\otimes |0\rangle\langle 0|\Big)\Big]+
\tr\Big[\rho_{2|3}\Big(\frac{\openone + \hat{Y}}{2}\otimes |0\rangle\langle 0|\Big)\Big]\Big\} \\
+(1-q)\Big\{
&\tr\Big[\rho_{1|1}\Big(\frac{\openone + \hat{Z}}{2}\otimes |1\rangle\langle 1|\Big)\Big]+
\tr\Big[\rho_{2|1}\Big(\frac{\openone - \hat{Z}}{2}\otimes |1\rangle\langle 1|\Big)\Big]
+\tr\Big[\rho_{1|2}\Big(\frac{\openone + \hat{X}}{2}\otimes |1\rangle\langle 1|\Big)\Big]\\
+&\tr\Big[\rho_{2|2}\Big(\frac{\openone - \hat{X}}{2}\otimes |1\rangle\langle 1|\Big)\Big]
+\tr\Big[\rho_{1|3}\Big(\frac{\openone + \hat{Y}}{2}\otimes |1\rangle\langle 1|\Big)\Big]+
\tr\Big[\rho_{2|3}\Big(\frac{\openone - \hat{Y}}{2}\otimes |1\rangle\langle 1|\Big)\Big]
\Big\}\Big\}.
\end{aligned}
\label{EqApp_fidelity_EBI_before_DI}
\end{equation}
Using the substitutions: $\openone\otimes\hat{Z}=-iB_5^{**}B_6^{**}B_7^{**}$, $\hat{Y}\otimes\openone=-iB_5^{**}B_6^{**}$, $\hat{Z}\otimes\hat{Z}=-iB_6^{**}B_7^{**}$, $\hat{X}\otimes\hat{Z}=-iB_7^{**}B_5^{**}$
, we have
\begin{equation}
\begin{aligned}
&\mathcal{F}[\bm{\sigma}^{**},\Lambda(\bm{\rho})] =  \frac{1}{12}\cdot\frac{1}{q^2+(1-q)^2}\cdot
\Big\{q
\Big\{\\
&\tr\Big[\rho_{1|1}\Big(\openone + B_5^{**} - iB_5^{**}B_6^{**}B_7^{**} - iB_6^{**}B_7^{**}\Big)\Big]+
\tr\Big[\rho_{2|1}\Big(\openone - B_5^{**} - iB_5^{**}B_6^{**}B_7^{**} + iB_6^{**}B_7^{**}\Big)\Big]\\
+&\tr\Big[\rho_{1|2}\Big(\openone + B_6^{**} - iB_5^{**}B_6^{**}B_7^{**} - iB_7^{**}B_5^{**}\Big)\Big]
+\tr\Big[\rho_{2|2}\Big(\openone - B_6^{**} - iB_5^{**}B_6^{**}B_7^{**} + iB_7^{**}B_5^{**}\Big)\Big]\\
+&\tr\Big[\rho_{1|3}\Big(\openone - B_7^{**} - iB_5^{**}B_6^{**}B_7^{**} + iB_5^{**}B_6^{**}\Big)\Big]+
\tr\Big[\rho_{2|3}\Big(\openone + B_7^{**} - iB_5^{**}B_6^{**}B_7^{**} - iB_5^{**}B_6^{**}\Big)\Big]\Big\} \\
+&(1-q)\Big\{\\
&\tr\Big[\rho_{1|1}\Big(\openone + B_5^{**} + iB_5^{**}B_6^{**}B_7^{**} + iB_6^{**}B_7^{**}\Big)\Big]+
\tr\Big[\rho_{2|1}\Big(\openone - B_5^{**} + iB_5^{**}B_6^{**}B_7^{**} - iB_6^{**}B_7^{**}\Big)\Big]\\
+&\tr\Big[\rho_{1|2}\Big(\openone + B_6^{**} + iB_5^{**}B_6^{**}B_7^{**} + iB_7^{**}B_5^{**}\Big)\Big]
+\tr\Big[\rho_{2|2}\Big(\openone - B_6^{**} + iB_5^{**}B_6^{**}B_7^{**} - iB_7^{**}B_5^{**}\Big)\Big]\\
+&\tr\Big[\rho_{1|3}\Big(\openone - B_7^{**} + iB_5^{**}B_6^{**}B_7^{**} - iB_5^{**}B_6^{**}\Big)\Big]+
\tr\Big[\rho_{2|3}\Big(\openone + B_7^{**} + iB_5^{**}B_6^{**}B_7^{**} + iB_5^{**}B_6^{**}\Big)\Big]
\Big\}\Big\}.
\end{aligned}
\label{EqApp_DI_fidelity_EBI}
\end{equation}
When we choose $\bm{\rho}=\bm{\sigma}^{**}$, our numerical result\footnote{{\bluv For each $a$ and $x$, we uniformly generate $10^5$ points between $q=0$ and $q=1$. Accordingly, we have $10^5$ values of $-i\tr(\sigma_{a|x}^{**}B_5^{**}B_6^{**}B_7^{**})$. We find that all of them are equal to $q-1/2$, up to the numerical precision of $10^{-17}$ (see \texttt{relations{\_}Eq{\_}B12.m} in \cite{SLChen_code} for the numerical results). The numerical check is the same for Eqs.~\eqref{EqApp_numerical_results} and \eqref{EqApp_SubHalf}.}} shows that $-i\tr(\rho_{a|x}B_5^{**}B_6^{**}B_7^{**})=(q-1/2)$ for all $a,x$,
\begin{equation}
-i\tr(\rho_{a|x}B_k^{**}B_l^{**})= \left\{
\begin{matrix}
(-1)^{a-1}(q-1/2) \quad\text{for}~~(x,k,l)=(1,6,7),(2,7,5),\\~\\
(-1)^{a}(q-1/2) \quad\text{for}~~(x,k,l)=(3,5,6),
\end{matrix}\right.
\label{EqApp_numerical_results}
\end{equation}
and
\clearpage
\begin{equation}
\frac{1}{2}=\left\{
\begin{matrix}
(-1)^{a-1}\tr(\rho_{a|x}B_j^{**}) \quad\text{for}~~(x,j)=(1,5),(2,6),\\~\\
(-1)^{a}\tr(\rho_{a|x}B_j^{**}) \quad\text{for}~~(x,j)=(3,7),
\end{matrix}\right.
\label{EqApp_SubHalf}
\end{equation}
which motivates us to use the following substitutions:
\begin{equation}
-i\tr(\rho_{a|x}B_5^{**}B_6^{**}B_7^{**})= \left\{
\begin{matrix}
(-1)^{a-1}(2q-1)\tr(\rho_{a|x}B_j^{**})\quad\text{for}~~x=1,2\\~\\
(-1)^{a}(2q-1)\tr(\rho_{a|x}B_j^{**})\quad\text{for}~~x=3
\end{matrix}\right.
\end{equation}
and
\begin{equation}
i\tr(\rho_{a|x}B_k^{**}B_l^{**})=-(2q-1)\tr(\rho_{a|x}B_j^{**})\quad\forall~a,
\end{equation}
where $(x,j,k,l)=(1,5,6,7),(2,6,7,5),(3,7,5,6)$. {\bluiv Gathering all above together, we obtain the following fidelity in terms of the optimal observables and assemblage:
\begin{equation}
\begin{aligned}
\mathcal{F} = &\frac{1}{12}\cdot\frac{1}{q^2+(1-q)^2}\Big[ 3 + (8q^2-8q+3)\cdot\\
&\big[\tr(\rho_{1|1}B_5^{**})-\tr(\rho_{2|1}B_5^{**}) + \tr(\rho_{1|2}B_6^{**})-\tr(\rho_{2|2}B_6^{**}) + \tr(\rho_{2|3}B_7^{**})-\tr(\rho_{1|3}B_7^{**}) \big]\Big].
\end{aligned}
\label{EqApp_DI_F_EBI_0}
\end{equation}
Now, using Eq.~\eqref{EqApp_SubHalf}, we can substitute the first term in the bracket of the above equation, namely, we have $3=\tr(\rho_{1|1}B_5^{**})-\tr(\rho_{2|1}B_5^{**}) + \tr(\rho_{1|2}B_6^{**})-\tr(\rho_{2|2}B_6^{**}) + \tr(\rho_{2|3}B_7^{**})-\tr(\rho_{1|3}B_7^{**})$. By such, we obtain the following fidelity which is independent of $q$:
\begin{equation}
\mathcal{F} = \frac{1}{3}\Big[\tr(\rho_{1|1}B_5^{**})-\tr(\rho_{2|1}B_5^{**}) + \tr(\rho_{1|2}B_6^{**})-\tr(\rho_{2|2}B_6^{**}) + \tr(\rho_{2|3}B_7^{**})-\tr(\rho_{1|3}B_7^{**}) \Big].
\label{EqApp_DI_F_EBI}
\end{equation}

Finally, dropping all the assumptions on $\rho_{a|x}$ and $B_j^{**}$, we have $3\geq\tr(\rho_{1|1}B_5^{})-\tr(\rho_{2|1}B_5^{}) + \tr(\rho_{1|2}B_6^{})-\tr(\rho_{2|2}B_6^{}) + \tr(\rho_{2|3}B_7^{})-\tr(\rho_{1|3}B_7^{})$. Therefore, a DI lower bound on the fidelity $\mathcal{F}[\bm{\sigma}^{**},\Lambda(\bm{\rho})]$ is obtained, namely,  $\mathcal{F} \geq \frac{1}{3}\Big[\tr(\rho_{1|1}B_5^{})-\tr(\rho_{2|1}B_5^{}) + \tr(\rho_{1|2}B_6^{})-\tr(\rho_{2|2}B_6^{}) + \tr(\rho_{2|3}B_7^{})-\tr(\rho_{1|3}B_7^{}) \Big]$. Together with Eq.~\eqref{EqApp_PolarDecomp}, we arrive the SDP form of Eq.~\eqref{Eq_min_F_SDP_EBI}.
}

\subsection{The $I_{3622}$ scenario}
Like the situation in self-testing of states and measurements, it is also possible to self-test the same assemblage with different Bell inequalities. For instance, consider the Bell inequality proposed by Ac{\'i}n \emph{et al.}~\cite{Acin16}:
\begin{equation}
\begin{aligned}
I_{3622}:=
&\langle A_1B_1\rangle + \langle A_1B_2\rangle + \langle A_2B_1\rangle - \langle A_2B_2\rangle\\
+&\langle A_1B_3\rangle + \langle A_1B_4\rangle - \langle A_3B_3\rangle + \langle A_3B_4\rangle\\
+&\langle A_2B_5\rangle + \langle A_2B_6\rangle - \langle A_3B_5\rangle + \langle A_3B_6\rangle\stackrel{\mathcal{L}}{\leq} 6,
\end{aligned}
\label{Eq_I3622}
\end{equation}
It was shown~\cite{Acin16} that the maximal quantum violation of $I_{3622}$, i.e., $6\sqrt{2}\approx 8.4853$, can be achieved if the shared state is the maximally entangled state and Alice's observables are the three Pauli observables. Such a strategy is the same as the elegant Bell scenario, which means that the reference assemblages for the both scenarios are the same. Consequently, the derivation of the SDP computing lower bound on the fidelity in the $I_{3622}$ scenario is almost the same as previous subsection. The difference is that Bob has $6$ measurement settings here and the optimal observables are $B_y^{**}=B_y^*\otimes |0\rangle\langle 0| + (B_y^*)^\mathsf{T}\otimes |1\rangle\langle 1|$, with
\begin{equation}
\begin{aligned}
B_1^*=\frac{\hat{Z}+\hat{X}}{\sqrt{2}},\quad B_2^*=\frac{\hat{Z}-\hat{X}}{\sqrt{2}},\quad B_3^*=\frac{\hat{Z}+\hat{Y}}{\sqrt{2}},\quad
B_4^*=\frac{\hat{Z}-\hat{Y}}{\sqrt{2}},\quad B_5^*=\frac{\hat{X}+\hat{Y}}{\sqrt{2}},\quad B_6^*=\frac{\hat{X}-\hat{Y}}{\sqrt{2}}.
\end{aligned}
\end{equation}
Therefore we introduce observables $B_7^{**}$, $B_8^{**}$, and $B_9^{**}$, such that (cf. Eq.~\eqref{EqApp_PolarDecomp})
\begin{equation}
\begin{aligned}
&B_7^{**}\left[\frac{1}{\sqrt{2}}\Big(B_1^{**}+B_2^{**}\Big)\right]\succeq 0,\quad
&B_8^{**}\left[\frac{1}{\sqrt{2}}\Big(B_5^{**}+B_6^{**}\Big)\right]\succeq 0,\quad
&B_9^{**}\left[\frac{1}{\sqrt{2}}\Big(B_3^{**}-B_4^{**}\Big)\right]\succeq 0,\\
&B_7^{**}\left[\frac{1}{\sqrt{2}}\Big(B_3^{**}+B_4^{**}\Big)\right]\succeq 0,\quad
&B_8^{**}\left[\frac{1}{\sqrt{2}}\Big(B_1^{**}-B_2^{**}\Big)\right]\succeq 0,\quad
&B_9^{**}\left[\frac{1}{\sqrt{2}}\Big(B_5^{**}-B_6^{**}\Big)\right]\succeq 0.\\
\end{aligned}
\label{EqApp_PolarDecomp2}
\end{equation}
The roles that $B_7^{**},B_8^{**},B_9^{**}$ play are the same as $B_5^{**},B_6^{**},B_7^{**}$ in the elegant Bell scenario. Following the procedure in the previous subsection, lower bounds on the fidelity can be computed with the SDP below:
\begin{equation}
\begin{aligned}
\min\quad &\mathcal{F}[\bm{\sigma}^{**},\Lambda(\bm{\rho})]\\
\text{s.t.} \quad &I_{3622}(\mathbf{P}) = I_{3622}^\text{obs},\quad \chi[\sax,\mathcal{S}]\succeq 0,\quad P_{\bm{\rho}}(a|x)=P_{\bm{\rho}}^{\rm obs}(a|x),\\
& \sum_a\chi[\sax,\mathcal{S}] = \sum_a\chi[\sigma_{a|x'},\mathcal{S}],\\
&\chi_\text{\tiny L}[\sax,B_7(B_1+B_2),\mathcal{S}']\succeq 0,\quad
\chi_\text{\tiny L}[\sax,B_8(B_5+B_6),\mathcal{S}']\succeq 0,\quad
\chi_\text{\tiny L}[\sax,B_9(B_3-B_4),\mathcal{S}']\succeq 0,\\
&\chi_\text{\tiny L}[\sax,B_7(B_3+B_4),\mathcal{S}']\succeq 0,\quad
\chi_\text{\tiny L}[\sax,B_8(B_1-B_2),\mathcal{S}']\succeq 0,\quad
\chi_\text{\tiny L}[\sax,B_9(B_5-B_6),\mathcal{S}']\succeq 0.
\end{aligned}
\label{EqApp_min_F_SDP_I3622}
\end{equation}
The result is plotted in Fig.~\ref{Fig_DI_fidelity_I3622_contr_mix}. 

\begin{figure*}[h!]
\centering
\includegraphics[width=8.5cm]{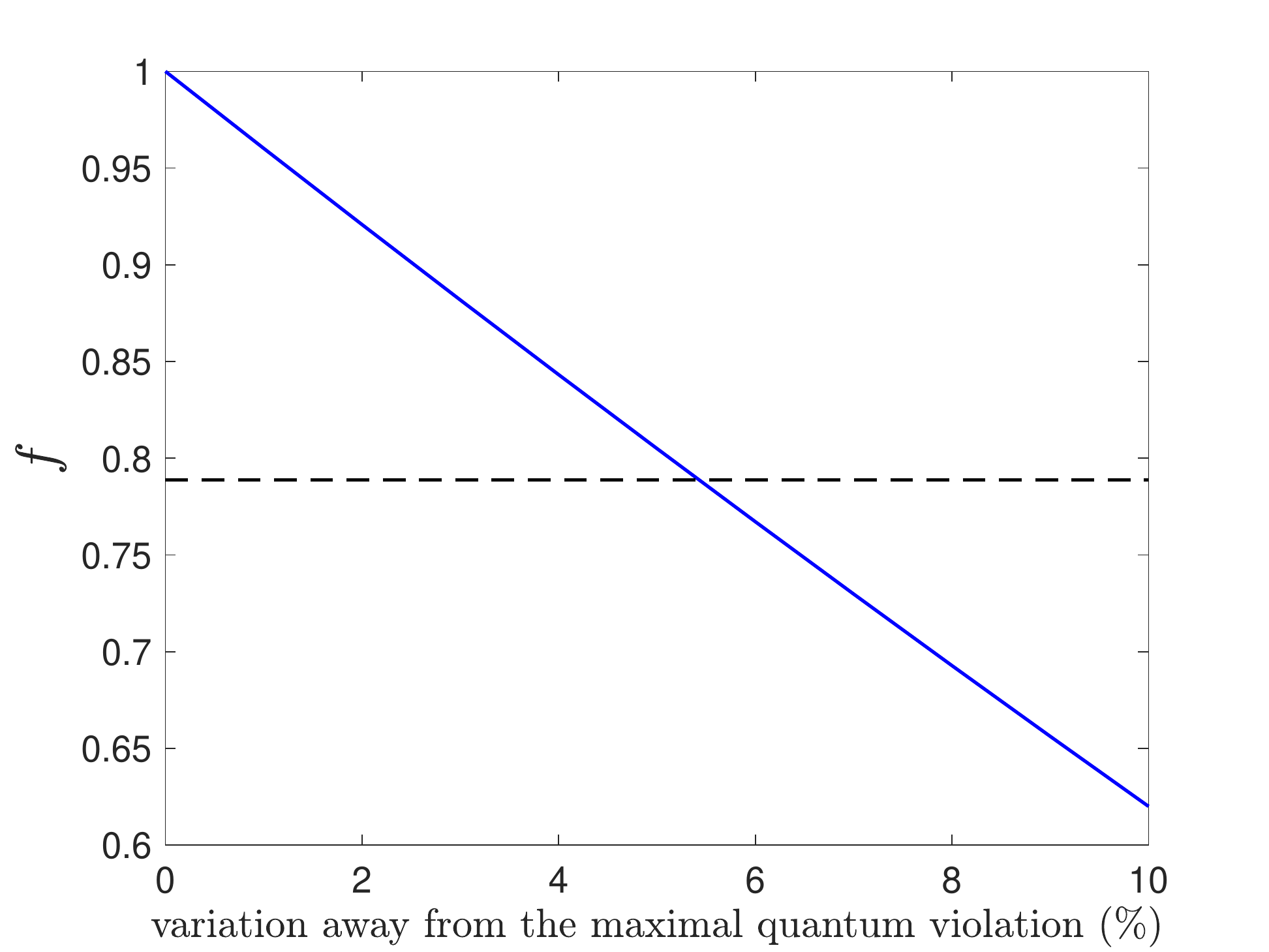}\\
\caption{
Robust self-testing of the $I_{3622}$ assemblage. The sequences of moments used for carry out the computation is attached in Appendix \ref{SecApp_Sequence}.
}
\label{Fig_DI_fidelity_I3622_contr_mix}
\end{figure*}


}

{\bluiv
\subsection{Remark}
Here we would like to add a remark on the DI expression of the fidelity.  Readers may consider Eq.~\eqref{EqApp_DI_F_EBI_0} as the final expression of the fidelity and relax it into the DI formulation.  In that case,  given the maximal quantum violation of the Bell inequality and given a value of $q$,  one can still obtain the fidelity with the value of $1$. This seems to contradict the fact that one cannot distinguish different ratios $q$ of the controlled mixture $q\bm{\sigma}^*\otimes\ket{0}\bra{0}+(1-q)(\bm{\sigma}^{*})^\mathsf{T}\otimes\ket{1}\bra{1}$ in a DI scenario as all the assemblages give the same behavior $P(a,b|x,y)$ (i.e.,  like the case of Eq.~\eqref{Eq_invariance}). {\bluv The key point is that if we assign a fixed value on $q$, the entire framework will not be DI anymore. Namely, the final expression of the fidelity is obtained by making assumption on $q$ therefore the result will not be a self-testing statement. In an extreme case, if we further assume that the CPTP map $\Lambda$ is equal to the identity map, yielding $\rho_{a|x}$ and $B_y$ to be optimal, then the final expression of the fidelity will be exactly equal to $1$. Apparently, such a result is even further away from a self-testing statement.}

}

\clearpage
\section{Related detail of applications on DI certification of all two-qubit entangled states}\label{SecApp_DIEW}
\subsection{Positivity of DI entanglement witnesses for separable states (the perfect self-testing)}
 
In this section we will prove that the DI entanglement witness $I_{\rm DIEW}$ can be used for certifying all entangled two-qubit states if both self-testing of Alice's and Bob's assemblages (prepared by Charlie and Daisy, respectively) are perfect. To this end, we need some ingredients for the proof. Recall from Definition~\ref{Def_1} in the main text, we have the relation $\Lambda(\bm{\rho})=\bm{\sigma}^*$ for the perfect self-testing scenario. From the Stinespring dilation~\cite{Stinespring55}, the action of the CPTP map $\Lambda$ on $\bm{\rho}$ can be treated as attaching an auxiliary state followed by an unitary operation, then tracing out the related system. Namely,
\begin{equation}
U(\bm{\rho}\otimes\varphi)U^\dag = \xi\otimes\bm{\sigma}^*,
\end{equation}
where $\varphi\in\mathsf{L}(\mathcal{H}_{\text{A}_0'})$ and $\xi\in\mathsf{L}(\mathcal{H}_{\text{A}_0})$ are some auxiliary states and $U$ is an unitary operation. Like the protocol of the standard self-testing of quantum states, here we also assume the composite system $U(\bm{\rho}\otimes\varphi)U^\dag$ is a tensor product state (i.e., the form of r.h.s. of the above equation) since each reference assemblages $\bm{\sigma}^*$ considered in this work is pure, i.e., it can not be written as a convex combination of other two assemblages. Rewriting the above equation, we have $\bm{\rho} =\tr_{\text{A}_0'} \big[ U^\dag (\xi\otimes\bm{\sigma}^*)U\big]:=\mathcal{E}(\bm{\sigma}^*)$, where the map $\mathcal{E}: \mathsf{L}(\mathcal{H}_{\text{A}'_0})\rightarrow\mathsf{L}(\mathcal{H}_{\text{A}_0})$ is CPTP according to the Stinespring dilation~\cite{Stinespring55}. In summary, for the perfect self-testing of real-valued assemblages, we have
\begin{equation}
\Lambda(\bm{\rho})=\bm{\sigma}^* \quad {\rm and} \quad \mathcal{E}(\bm{\sigma}^*)=\bm{\rho}.
\end{equation}
{\blubb
Similarly, for the perfect self-testing of complex-valued assemblages (cf. Definition~\ref{Def_2}), we have
\begin{equation}
\Lambda(\bm{\rho})=q\bm{\sigma}^*\otimes |0\rangle\langle 0| + (1-q)(\bm{\sigma}^*)^\mathsf{T}\otimes |1\rangle\langle 1|
\quad{\rm and}\quad
\mathcal{E}\Big(q\bm{\sigma}^*\otimes |0\rangle\langle 0| + (1-q)(\bm{\sigma}^*)^\mathsf{T}\otimes |1\rangle\langle 1|\Big)=\bm{\rho}\quad
\end{equation}
for some CPTP maps $\Lambda$ and $\mathcal{E}$.

}

Now we return to the entanglement certification protocol. Recall that the DI entanglement witness is written as (cf. Eq.~\eqref{Eq_DIEW})
\begin{equation}
I_{\rm DIEW}:=\sum_{c,d,u,v}\beta_{c,d}^{u,v} P(c,+,+,d|u,\blacklozenge,\blacklozenge,v),
\end{equation}
where $x=\blacklozenge$ and $y=\blacklozenge$, respectively, represent for Alice's and Bob's $7$th measurement settings. In quantum theory, we denote the $7$th settings as POVMs $\{\EAA,\openone-\EAA\}$ and $\{\EBB,\openone-\EBB\}$, thus $a=+$ and $b=+$ in the above equation are outcomes corresponding to $\EAA$ and $\EBB$, respectively. With this, the quantum realization of $P(c,+,+,d|u,\blacklozenge,\blacklozenge,v)$ is given by
\begin{equation}
\begin{aligned}
P(c,+,+,d|u,\blacklozenge,\blacklozenge,v) &= \tr\big[(E_{c|u}^{\text{C}}\otimes\EAA\otimes\EBB\otimes E_{d|v}^{\text{D}})(\rca\otimes\rab\otimes\rbd)\big]\\
&=\tr\big[ (\EAA\otimes\EBB)(\tcu\otimes\rab\otimes\odv) \big].
\end{aligned}
\end{equation}

If Alice's and Bob's reference assembalges, i.e., $\{\tcustar\}$ and $\{\odvstar\}$, are both tomographically complete, they can be used to span an Hermitian observable, including the entanglement witness $W^*$~\cite{Branciard13,Bowles18,Bowles18PRA}:
\begin{equation}
\begin{aligned}
W^* = \sum_{c,d,u,v}\beta_{c,d}^{u,v}(\tcustar)^\mathsf{T}\otimes(\odvstar)^\mathsf{T}
\end{aligned}
\end{equation}
Then, for any separable state $\rab=\sum_k p_k\hat{\sigma}_k^\text{A}\otimes\hat{\sigma}_k^\text{B}$ we have
{\blubb
\begin{equation}
\begin{aligned}
&I_{\rm DIEW}\\
&=\sum_{c,d,u,v}\beta_{c,d}^{u,v}\tr\big[ (\EAA\otimes\EBB)(\tcu\otimes\rab\otimes\odv) \big]\\
&=\sum_{c,d,u,v,k}p_k\beta_{c,d}^{u,v}\tr\Big\{ (E_k^{\text{A}_0}\otimes E_k^{\text{B}_0})\\
&\quad\quad \Big[
\mathcal{E}_{\bm{\tau}}\Big(q_1\tcustar\otimes|0\rangle\langle 0|+(1-q_1)(\tcustar)^\mathsf{T}\otimes|1\rangle\langle 1|\Big)\otimes
\mathcal{E}_{\bm{\omega}}\Big(q_2\odvstar\otimes|0\rangle\langle 0|+(1-q_2)(\odvstar)^\mathsf{T}\otimes|1\rangle\langle 1|\Big) \Big] \Big\}\\
&=\sum_{c,d,u,v,k}p_k\beta_{c,d}^{u,v}\Big\{
q_1q_2\tr\big[ (F_k^{\text{A}'_0}\otimes F_k^{\text{B}'_0})(\tcustar\otimes\odvstar)\big] +
q_1(1-q_2)\tr\big[ (F_k^{\text{A}'_0}\otimes \tilde{F}_k^{\text{B}'_0})(\tcustar\otimes(\odvstar)^\mathsf{T})\big] +\\
&\quad\quad (1-q_1)q_2\tr\big[ (\tilde{F}_k^{\text{A}'_0}\otimes F_k^{\text{B}'_0})((\tcustar)^\mathsf{T}\otimes\odvstar)\big] +
(1-q_1)(1-q_2)\tr\big[ (\tilde{F}_k^{\text{A}'_0}\otimes \tilde{F}_k^{\text{B}'_0})((\tcustar)^\mathsf{T}\otimes(\odvstar)^\mathsf{T})\big]
\Big\}\\
&=\sum_k p_k \Big\{ q_1q_2\tr\big[ W^* (F_k^{\text{A}'_0})^\mathsf{T} \otimes (F_k^{\text{B}'_0})^\mathsf{T} \big]+
q_1(1-q_2)\tr\big[ W^* (F_k^{\text{A}'_0})^\mathsf{T} \otimes \tilde{F}_k^{\text{B}'_0}\big]+\\
&\quad\quad (1-q_1)q_2\tr\big[ W^* \tilde{F}_k^{\text{A}'_0}\otimes (F_k^{\text{B}'_0})^\mathsf{T}\big]  + 
(1-q_1)(1-q_2)\tr\big[ W^* \tilde{F}_k^{\text{A}'_0}\otimes \tilde{F}_k^{\text{B}'_0} \big]\Big\}
\geq 0,
\end{aligned}
\label{Eq_App_positive_IDIEW}
\end{equation}
where we define $E_k^{\text{A}_0}:=\tr_{\text{A}}[\EAA (\openone\otimes\hat{\sigma}_k^\text{A})] $ and $E_k^{\text{B}_0}:=\tr_{\text{B}}[\EBB (\hat{\sigma}_k^\text{B}\otimes \openone)] $ in the second equality. If we consider $\tcustar\otimes|0\rangle\langle0|$ as a result of a CP map $\mathcal{E}_0(\cdot):=(\cdot)\otimes|0\rangle\langle 0|$, then we have $\tr[E_k^{\text{A}_0}\mathcal{E}_{\bm{\tau}}\circ\mathcal{E}_0 (\tcustar)]=\tr[\mathcal{E}_0^\dag\circ\mathcal{E}_{\bm{\tau}}^\dag(E_k^{\text{A}_0})\tcustar]:=\tr(F_k^{\text{A}'_0}\tcustar)$, where $\mathcal{E}^\dag$ is the dual of $\mathcal{E}$, which is still CP. Similarly, $\tilde{F}_k^{\text{A}'_0}:=\mathcal{E}_1^\dag\circ\mathcal{E}_{\bm{\tau}}^\dag(E_k^{\text{A}_0})$ with $\mathcal{E}_1(\cdot):=(\cdot)\otimes|1\rangle\langle 1|$ (the similar notations for the part of $\odvstar$).
}
The inequality holds since $W^*$ is an entanglement witness, therefore $\tr(W^* O)$ is non-negative for all separable positive semidefinite operators $O$.

{\blubb To show that there exists a quantum strategy for $I_{\rm DIEW}$ to detect the two-qubit entangled state $\rab$, we choose $\EAA$ and $\EBB$ as the projection onto the maximally entangled state, i.e., $\EAA=\EBB=\ket{\Phi^+}\bra{\Phi^+}$ with $\ket{\Phi^+}:=(1/\sqrt{2})(\ket{00}+\ket{11})$, and $\tcu$, $\odv$ are chosen as $\tcustar$, $\odvstar$, i.e., the elegant Bell assemblage described in Eq.~\eqref{Eq_EB_assemblage}. Thus we obtain
\begin{equation}
\begin{aligned}
I_{\rm DIEW}
&=\sum_{c,d,u,v}\beta_{c,d}^{u,v}\tr\big[ (\EAA\otimes\EBB)(\tcu\otimes\rab\otimes\odv) \big]\\
&=\sum_{c,d,u,v}\beta_{c,d}^{u,v}\tr\Big[ \left(\ket{\Phi^+}\bra{\Phi^+}\otimes\ket{\Phi^+}\bra{\Phi^+}\right)\left(\tcustar\otimes\rab\otimes\odvstar\right) \Big]\\
&=\frac{1}{4}\sum_{c,d,u,v}\beta_{c,d}^{u,v}\tr\left\{ (\tcustar)^\mathsf{T}\otimes(\odvstar)^\mathsf{T} \rab \right\}\\
&= \frac{1}{4}\tr(W^*\rab) < 0.
\end{aligned}
\end{equation}

}

\subsection{Shift of the separable bound of DI entanglement witness (imperfect self-testing)}
{\blubb
Following the previous section, for the case where the quantum violation of $I_{3622}$ departs from the maximal value of $6\sqrt{2}$, one has the inequivalence between $\tcu$ and $\mathcal{E}_{\bm{\tau}}(q_1\tcustar\otimes |0\rangle\langle 0| + (1-q_1)(\tcustar)^\mathsf{T}\otimes |1\rangle\langle 1|)$. Namely, $||\tcu-\mathcal{E}_{\bm{\tau}}(\tcustarstar)||\leq\eta$ for some positive number $\eta$, where we define $\tcustarstar:=q_1\tcustar\otimes |0\rangle\langle 0| + (1-q_1)(\tcustar)^\mathsf{T}\otimes |1\rangle\langle 1|$. Here, we would like to show the separable bound of $I_{3622}$ is shifted by
\begin{equation}
I_{\rm DIEW}\geq-r(\eta),
\label{Eq_DIEW_sep_bound}
\end{equation}
with $r(\eta)$ being some positive function satisfying the condition: $r(\eta)= 0$ when $\eta= 0$.
\begin{proof}
The proof follows the technique of Ref.~\cite{Bowles18PRA}. The relation $||\tcu-\mathcal{E}_{\bm{\tau}}(\tcustarstar)||\leq\eta$ can be reformulated as: 
\begin{equation}
\tcu = \mathcal{E}_{\bm{\tau}}(\tcustarstar) + \Delta_{c|u}
\end{equation}
with $||\Delta_{c|u}||\leq\eta$. Note that $\eta=0$ implies the perfect self-testing. For Bob's assemblage, one can obtain the similar relation: $\odv=\mathcal{E}_{\bm{\omega}}(\odvstarstar) + \Delta_{d|v}$ with $||\Delta_{d|v}||\leq\eta$. Then, for a separable state $\rab=\sum_k p_k \hat{\sigma}_k^{\text{A}}\otimes\hat{\sigma}_k^{\text{B}}$, the value of $I_{\rm DIEW}$ is
\begin{equation}
\begin{aligned}
I_{\rm DIEW}
&=\sum_{c,d,u,v}\beta_{c,d}^{u,v}\tr\big[ (\EAA\otimes\EBB)(\tcu\otimes\rab\otimes\odv) \big]\\
&=\sum_{c,d,u,v,k} p_k \beta_{c,d}^{u,v}\tr\left\{ (\EAA\otimes\EBB)\Big[\big[ \mathcal{E}_{\bm{\tau}}(\tcustarstar) + \Delta_{c|u} \big]\otimes\hat{\sigma}_k^{\text{A}}\otimes\hat{\sigma}_k^{\text{B}}\otimes \big[ \mathcal{E}_{\bm{\omega}}(\odvstarstar) + \Delta_{d|v}\big] \Big]\right\}\\
&=I_{\rm DIEW}^{\rm noiseless} + \sum_{c,d,u,v,k} p_k \beta_{c,d}^{u,v}\Big[
\tr\big( \EAA~\mathcal{E}_{\bm{\tau}}(\tcustarstar)\otimes\sigmakA \big)\tr\big( \EBB~\sigmakB\otimes\Ddv\big)\\
&\quad+\tr\big( \EAA~\Dcu\otimes\sigmakA \big)\tr\big( \EBB~\sigmakB\otimes\mathcal{E}_{\bm{\omega}}(\odvstarstar)\big)
+\tr\big( \EAA~\Dcu\otimes\sigmakA \big)\tr\big( \EBB~\sigmakB\otimes\Ddv\big)\Big], 
\end{aligned}
\end{equation}
where $I_{\rm DIEW}^{\rm noiseless}$ is the DI entanglement wintess $I_{\rm DIEW}$ for $\eta=0$. As considered in Ref.~\cite{Bowles18PRA}, under the worst-case scenario, one has $I_{\rm DIEW}^{\rm noiseless}=0$ and that all the terms in the summation give negative numbers. With the bound relations~\cite{Bowles18PRA}
\begin{equation}
\begin{aligned}
& |\tr\big( \EAA~\mathcal{E}_{\bm{\tau}}(\tcustarstar)\otimes\sigmakA \big)|\leq||\mathcal{E}_{\bm{\tau}}(\tcustarstar)||\leq||\tcustarstar||=\frac{1}{2},\\
&|\tr\big( \EAA~\Dcu\otimes\sigmakA \big)|\leq||\Dcu||\leq\eta,
\end{aligned}
\end{equation}
we can obtain a lower bound on $I_{\rm DIEW}$, which is a function of $\eta$, where $I_{\rm DIEW}\geq0$ when $\eta= 0$, arriving the statement of Eq.~\eqref{Eq_DIEW_sep_bound}.

\end{proof}

}

\section{DI certification of all non entanglement-breaking qubit channels}\label{SecApp_DI_nonEBC}
In this section, we provide the detail proof of the faithfulness of the DI witnesses of non entanglement-breaking (non-EB) qubit channels of Eq.~\eqref{Eq_DI_nonEBC}. That is, given any non-EB qubit channel, there exists a witness $\{\gamma_{c,d}^{u,v}\}$, namely a set of coefficients, such that
\begin{equation}
I_{\mathcal{N}}:=\sum_{c,d,u,v}\gamma_{c,d}^{u,v} P_{\mathcal{N}}(+,d|c,u,\blacklozenge,v) < 0
\label{EqApp_P_nonEB_DI}
\end{equation}
while $\sum_{c,d,u,v}\gamma_{c,d}^{u,v} P_{\mathcal{N}}^{\rm EB}(+,d|c,u,\blacklozenge,v)\geq 0$ for all EB channels.
\begin{proof}
Recall that for a non-EB channel $\mathcal{N}$, its CJ matrix $J_\mathcal{N}:=(\mathcal{N}\otimes\openone)(\ket{\Phi^+}\bra{\Phi^+})\in \mathsf{L}(\mathcal{H}_{\rm A})\otimes\mathsf{L}(\mathcal{H}_{\rm B})$ must be entangled~\cite{Horodecki03}, where $\ket{\Phi^+}$ is the maximally entangled state\footnote{Note that the usage of CJ matrix here is not the same as that of the CJ matrix used for relaxing the fidelity expression in Sec.~\ref{Sec_robust_ST}. }. This implies that there exists an entanglement witness $W$ such that $\tr(W J_\mathcal{N}) < 0$ while $\tr(W \rho) \geq 0$ for all $\rho$ separable on $\mathsf{L}(\mathcal{H}_{\rm A})\otimes\mathsf{L}(\mathcal{H}_{\rm B})$.

For the reference assemblages $\{\tcustar\}$ and $\{\odvstar\}$ being tomographically complete, they can span the witness $W$. For the later use, we use $\hattcustar$ (i.e., the normalized states of $\tcustar$) and $\odvstar$ to span $W$:
\begin{equation}
W = \sum_{c,d,u,v}\gamma_{c,d}^{u,v}~ (\hattcustar)^\mathsf{T} \otimes (\odvstar)^\mathsf{T}.
\end{equation}
In Eq.~\eqref{EqApp_P_nonEB_DI}, the quantum realization of $P_{\mathcal{N}}(+,d|c,u,\blacklozenge,v)$ is
\begin{equation}
P_{\mathcal{N}}(+,d|c,u,\blacklozenge,v) = \tr\Big[ \left(\EBB\otimes E_{d|v}^{\text{D}}\right)\left( \mathcal{N}(\hattcu)\otimes \rbd \right)\Big],
\end{equation}
where 1) $\EBB$ is the projection corresponding to the outcome $b=+$ when performing the measurement $\blacklozenge$, 2) $E_{d|v}^{\text{D}}$ are Daisy's POVM elements corresponding to measurement outcomes $d$ and inputs $v$, 3) and the channel $\mathcal{N}$ maps operators from $\mathsf{L}(\mathcal{H}_{\text{A}'_0})$ to $\mathsf{L}(\mathcal{H}_\text{B})$ (see Fig.~\ref{Fig_signaling_game_DI} in the main text for the overall DI setting). If $\mathcal{N}$ is an EB channel, the witness $I_{\mathcal{N}}$ is then written as
{\blubb
\begin{equation}
\begin{aligned}
I_{\mathcal{N}}&=\sum_{c,d,u,v}\gamma_{c,d}^{u,v}  \tr\Big[ \left(\EBB\otimes E_{d|v}^{\text{D}}\right)\left( \mathcal{N}(\hattcu)\otimes \rbd \right)\Big]\\
&=\sum_{c,d,u,v}\gamma_{c,d}^{u,v} \tr\Big[ \EBB \left(\mathcal{N}(\hattcu)\otimes\odv\right) \Big]\\
&=\sum_{c,d,u,v}\gamma_{c,d}^{u,v} \tr \Big\{ \EBB \Big[ \sum_\lambda\pi(\lambda)\sum_k\hat{\xi}_{k,\lambda}\tr(\Pi_{k|\lambda}\hattcu)  \Big] \otimes \odv \Big\}\\
&=\sum_{c,d,u,v,\lambda,k}\gamma_{c,d}^{u,v}\pi(\lambda) \tr(\Pi_{k|\lambda}\hattcu)\tr\big[\EBB(\hat{\xi}_{k,\lambda}\otimes\odv)\big]\\
&=\sum_{c,d,u,v,\lambda,k}\gamma_{c,d}^{u,v}\pi(\lambda) \tr\left\{ \Pi_{k|\lambda}\cdot \mathcal{E}_{\bm{\tau}}\Big(q_1\hattcustar\otimes |0\rangle\langle 0| + (1-q_1)(\hattcustar)^{\mathsf{T}}\otimes |1\rangle\langle 1|\Big) \right\} \\
&\quad\quad\tr\left\{ (E_{k,\lambda}\cdot \mathcal{E}_{\bm{\omega}}\Big(q_2\odvstar\otimes |0\rangle\langle 0| + (1-q_2)(\odvstar)^{\mathsf{T}}\otimes |1\rangle\langle 1|\Big) \right\}\\
\end{aligned}
\end{equation}
In the thrid equality we use the property of an EB channel, where $\hat{\xi}_{k,\lambda}$, $\Pi_{k|\lambda}$, and $\pi(\lambda)$ are, respectively, some quantum states, POVM elements, and probabilities~\cite{Rosset18}. In the last line, $E_{k,\lambda}:=\tr_{\text{B}}[\EBB(\hat{\xi}_{k,\lambda}\otimes\openone)]$. Following the same reasoning in Eq.~\eqref{Eq_App_positive_IDIEW}, the last line is a non-negative value.

To show that there exists a quantum strategy violating $I_{\mathcal{N}}\geq 0$ for the non-EB channel $\mathcal{N}$, we can choose $\hattcu$ as $\hattcustar$, $\odv$ as $\odvstar$, and $\EBB$ as the projection onto the maximally entangled state: $\EBB = \ket{\Phi^+}\bra{\Phi^+}$ with $\ket{\Phi^+}=(1/\sqrt{2})(\ket{00}+\ket{11})$. With these, $I_{\mathcal{N}}$ will be
\begin{equation}
\begin{aligned}
&\sum_{c,d,u,v}\gamma_{c,d}^{u,v}  \tr\Big[ \ket{\Phi^+}\bra{\Phi^+}\left( \mathcal{N}(\hattcustar)\otimes \odvstar \right)\Big]\\
=&\frac{1}{2}\sum_{c,d,u,v}\gamma_{c,d}^{u,v} \tr\Big[ \mathcal{N}(\hattcustar) \cdot (\odvstar)^\mathsf{T}   \Big]\\
=&\frac{1}{2}\sum_{c,d,u,v}\gamma_{c,d}^{u,v} \tr\Big[ J_{\mathcal{N}}\cdot (\hattcustar)^\mathsf{T} \otimes (\odvstar)^\mathsf{T}   \Big]\\
=&\frac{1}{2}\tr(J_{\mathcal{N}} W)< 0.
\end{aligned}
\end{equation}
In the first and second equality of the above equation, we use the relations $\bra{\Phi^+}A\otimes B\ket{\Phi^+}=\frac{1}{d}\tr(AB^\mathsf{T})$ and $\tr[\mathcal{N}(A)B]=\tr[J_\mathcal{N} (A^\mathsf{T}\otimes B)]$~\cite{WolfLect}.

}

\end{proof}

For the case of imperfect self-testing, the situation is similar to the previous section. The separable bound $0$ will be shifted and some non-EB qubit channels will be failed to detected. The technique of the proof is the same as the one in the previous section, therefore we do not repeat it here.

{\blu
\section{The sequences carrying out the computation}\label{SecApp_Sequence}
In this section, we explicitly write down the detail of the sequences (i.e., the set of operators) of the AMMs to carry out the computational results in the main text. For the CHSH scenario, the sequence $\mathcal{S}$ is in the $2$nd level, i.e.,
\begin{equation}
\mathcal{S} = \{\openone, B_1, B_2, B_1B_2, B_2B_1\}.
\end{equation}

For the tilted-CHSH scenario, $S$ is chosen as
\begin{equation}
\begin{aligned}
\mathcal{S} = \{&\openone, B_1, B_2, B_3, B_4,\\
     &B_1B_2, B_2B_1, B_4B_3, B_3B_4,B_3B_1, B_3B_2,\\
     &B_1B_3, B_2B_3,B_4B_1, B_4B_2,B_1B_4, B_2B_4,\\
     &B_1B_2B_1, B_2B_1B_2, B_3B_4B_3, B_4B_3B_4,\\
     &B_1B_3B_1, B_1B_3B_2, B_1B_4B_1, B_1B_4B_2,\\
     &B_1B_4B_3, B_1B_3B_4, B_2B_4B_3, B_2B_3B_4,
\end{aligned}
\end{equation}
of which the size is $29$. Meanwhile, the sequence $\mathcal{S'}$ of the localizing matrices is
\begin{equation}
\begin{aligned}
\mathcal{S'} = \{&\openone, B_1, B_2, B_3, B_4,\\
     &B_1B_2, B_2B_1, B_4B_3, B_3B_4,B_1B_4, B_4B_1\}.
\end{aligned}
\end{equation}

For the elegant-Bell scenario, we have $\mathcal{S}$ with the size of $38$:
\begin{equation}
\begin{aligned}
\mathcal{S} = \{&\openone, B_1, B_2, B_3, B_4, B_5, B_6, B_7,\\
     &B_5B_6, B_6B_5, B_5B_7, B_7B_5, B_6B_7, B_7B_6,\\
     &B_5B_1, B_5B_2, B_5B_3, B_5B_4,B_1B_5, B_2B_5, B_3B_5, B_4B_5,\\
     &B_6B_1, B_6B_2, B_6B_3, B_6B_4,B_1B_6, B_2B_6, B_3B_6, B_4B_6,\\
     &B_7B_1, B_7B_2, B_7B_3, B_7B_4,B_1B_7, B_2B_7, B_3B_7, B_4B_7\},
\end{aligned}
\end{equation}
while the sequence $\mathcal{S'}$ composing the localizing matrices is:
\begin{equation}
\begin{aligned}
\mathcal{S'} = \{&\openone, B_1, B_2, B_3, B_4, B_5, B_6, B_7\}
\end{aligned}
\end{equation}

Finally, for the $I_{3622}$ scenario, the size of $\mathcal{S}$ is 40:
\begin{equation}
\begin{aligned}
\mathcal{S} = \{&\openone, B_1, B_2, B_3, B_4, B_5, B_6, B_7, B_8, B_9,\\
     &B_7B_8, B_8B_7, B_7B_9, B_9B_7, B_8B_9, B_9B_8,\\
     &B_7B_1, B_7B_2, B_7B_3, B_7B_4,B_1B_7, B_2B_7, B_3B_7, B_4B_7,\\
     &B_8B_1, B_8B_2, B_8B_5, B_8B_6,B_1B_8, B_2B_8, B_5B_8, B_6B_8,\\
     &B_9B_3, B_9B_4, B_9B_5, B_9B_6,B_3B_9, B_4B_9, B_5B_9, B_6B_9\},
\end{aligned}
\end{equation}
while
\begin{equation}
\begin{aligned}
\mathcal{S'} = \{&\openone, B_1, B_2, B_3, B_4, B_5, B_6, B_7, B_8, B_9\}.
\end{aligned}
\end{equation}

}

\clearpage

\nocite{apsrev41Control}
\bibliographystyle{apsrev4-1}

%

\end{document}